\documentclass[pre,showpacs,reprint]{revtex4-1}
\usepackage{amsmath,amssymb}
\usepackage{rotating}
\usepackage[utf8]{inputenc}
\usepackage{graphicx} % Include figure files
\usepackage{epstopdf}

\begin{document}

\title{Simple pair-potentials and pseudo-potentials 
for warm-dense matter applications.}

\normalsize

\author
{
 M.W.C. Dharma-wardana}
\email[Email address:\ ]{chandre.dharma-wardana@nrc-cnrc.gc.ca}
\affiliation{
National Research Council of Canada, Ottawa, Canada, K1A 0R6
}

\date{\today}
\begin{abstract}
We present  computationally simple
 parameter-free pair potentials useful for solids, liquids
and plasma at arbitrary temperatures.
They successfully treat warm-dense matter (WDM) systems like
carbon or silicon with complex tetrahedral or other
structural bonding features.
Density functional theory  asserts that only one-body
electron densities, and one-body ion densities are needed
for a complete description of electron-ion systems. 
DFT is used here to reduce {\it both} the electron many-body problem
and the ion many-body problem to an exact one-body problem,
namely  that of the neutral pseudoatom (NPA). 
We compare the  Stillinger-Weber (SW) class of multi-center
potentials,  the embedded-atom approaches and $N$-atom DFT,
with the one-atom DFT approach of the NPA to
show that many-ion effects  are systematically included
in this one-center  method  via  one-body  exchange-correlation
functionals.  This computationally highly efficient
one-center  DFT-NPA approach is
contrasted with the usual $N$-center DFT calculations that are
coupled with molecular dynamics (MD) simulations to equilibriate
the ion distribution. Comparisons are given with the pair-potential
parts of the SW, `glue' models, and the corresponding NPA
pair-potentials to elucidate  how the NPA potentials capture
many-center effects using single-center one-body densities.   
\end{abstract}

\maketitle

\section{Introduction} Condensed-matter systems at temperatures $T\ge 0$,
 where the
constituents may be neutral or ionized, solid or fluid, are included
 in the appellation
`` warm dense matter'' (WDM). The reference to ``dense'' pertains mostly
to the property of strong interactions rather than to `density' {\it per se}. 
Furthermore, `warm' systems are those away from the simplifications available
for the $T\to 0$ or $T\to\infty$ limits. Hence a truly general finite-$T$
many-body theory of interacting electrons and ions in arbitrary
states of matter is needed as our focus is on first-principles methods.
Density functional theory (DFT) is a
favoured approach since it reduces
such problems to a theory of mere one-body densities, requiring
no multi-atom approaches, at least in principle.
Here we exploit DFT for simplifying the general electron-ion problem to the
fullest~\cite{ilciacco93, cdw-N-rep19}. 

While typical WDM systems are in the domain of plasma physics,
laser-matter interactions, high pressure physics, geophysics,
astrophysics etc., many problems in nano-structure physics also fall into
WDM physics. An electron layer or a hole layer in a quantum
well contains particles of modified effective mass ($m^*$) that may be a
tenth of the free electron mass. The corresponding Fermi energies and particle
densities are such that even at room temperature, the electrons may be partially
degenerate while the holes are classical particles, the system behaving as
warm-dense matter~\cite{lfc1-dw19}.
The quantum wells or nanostructures may have metal-oxide layers and
inhomogenieties that need  a unified many-body first
principles theory~\cite{2v2d04}. 

The need for practical computations 
 of energetics of defects and dislocations in metallic structures led to
semi-empirical theories usually known as embedded-atom or effective 
medium theories~\cite{DawFoilsBasks93}. The potential $V(\vec{R})$
felt by an ion at $\vec{R}$  is  modeled
using two-body, three-body and possibly higher terms in the energy,
and used to motivate functional forms for numerically fitted
representations of structure dependent  energies.

 A parallel
development  in semiconductor physics modified the usual two-body
potentials of `covalent bonds' (e.g., of the Lenard-Jones type)  with
 `structure dependent' three-body terms, viz., as in the well-known
 Stillinger-Weber~\cite{SW85} or Tersoff models~\cite{Ters88} for C, Si
 and other tetrahedral solids and fluids. These have been extended
to include empirical `bond-order', angular and `torsional'
 correlations~\cite{ghiring05}, producing very complex `potentials'.
Highly  parameter-loaded potentials fitted to
empirical data as well as large DFT data bases~\cite{OQMD} have
emerged. Many of these multi-center multi-parameter potentials have
been incorporated in codes like LAMMPS (Large-scale Atomic/Molecular
Massively Parallel Simulator)~\cite{LAMMPS95} for regular use in
molecular dynamics (MD) simulations. In
this approach, the electron system is fully integrated out
and only a classical simulation based on the multi center potentials is used.
Hence these multi-center effective-medium  models
lack  associated atomic pseudopotentials that may be used to delineate
 their building blocks, or for use in transport studies.
 In contrast, the NPA  potentials that we discuss here
relate directly to underlying
pseudopotentials, response functions, and XC-functionals as the NPA is
based on first-principles physics. It also includes the multi-center
`effective medium' effects via ion-ion XC-functionals
generated {\it in situ}.

Attempts to extend these multi-center potentials, and use $N$-center
DFT for $T>0$  applications to complex systems like, say,
WDM plastics~\cite{hamelCH12}, have exploited increasing
computer power without change in the conceptual framework.
However, such attempts have yielded disappointing results,
esp. with respect to WDM applications. Thus, SW and similar
potentials predict spurious phase transitions for liquid
carbon~\cite{glosli99, WuLPT02}, while the Tersoff potential
over-estimates the melting point of silicon by $\sim$ 50\%.
Similarly, Kraus {\it et al}~\cite{kraus13} found that sophisticated
`bond-order' potentials available for carbon failed badly for their
study of liquid-carbon at 100 GPa.

In the following we argue, and provide realistic examples
to show  that these empirical data-fitting approaches,  be they at the
level of Stillinger-Weber (SW)~\cite{SW85}, or  bond-order potentials, 
or with  ``high through-put data-base fittings''~\cite{OQMD},
are in fact not necessary,
at least for fluids and systems with
some simplifying symmetries, if the full power of DFT is exploited.
However, $N$-atom DFT is necessary and best used to provide
first-principles benchmarks for other methods including
one-atom DFT methods. 

If we consider an electron-ion system in equilibrium at
the temperature $T$,  with a one-body electron density $n(\vec{r})$,
and a one-body ion density $\rho(\vec{r})$,
standard DFT asserts that {\it all} the thermodynamic
properties and linear transport properties are a functional of just the
 one-body $n(r)$ and $\rho(r)$, while the many-body interactions are
also functionals of just these one body densities.
 Hence multi-center approaches should {\it not be necessary} if
DFT is rigorously applied.
The many-body effects are relegated to the  exchange-correlation
(XC) functionals. The functionals are non-local but remain
one-body density functionals. 
At finite-$T$ we use the free-energy functionals 
$F_{ss'}^{xc}[n_s,n_{s'}]$,
where $s$ or $s'=e,i$ is the species index, 
with $n_e=n(r), n_i=\rho(r)$~\cite{ilciacco93}.  
The finite-$T$ free-energy functionals
reduce to the usual ground-state energy functionals at $T=0$.

Common computer implementations of DFT in large codes like
ABINIT~\cite{ABINIT} or VASP~\cite{VASP} explicitly require the 
locations $\vec{R}_I, I=1,N$, of the nuclei of  $N$ ions and only the
electron problem is reduced by DFT. The density functional theory of
the one-body  ion density $\rho(r)$  is {\it not} invoked in these
codes, although the DFT for classical ions begun to be exploited since the
1970s, e.g., Evans~\cite{EvansDFT79}. Only an electron exchange-correlation
functional $F_{ee}^{xc}$ is  invoked in VASP or ABINIT,
using a Kohn-Sham (KS) calculation
for  specific ionic configuration ${\vec{R}_I}$ using the Born-Oppenheimer 
(BO) approximation. The ions merely provide an `external potential'. 
Classical molecular dynamics is used to equilibrate the ion distribution
$\sum \delta(\vec{r}-\vec{R_I})$. Such an approach produces a many-centered
 potential energy function $V({\vec{R}_1,\cdots,\vec{R}_N})$. So,
 single-ion properties, pair-potentials between ions etc., are not directly
 available by this $N$-center DFT, as the
calculation is for a solid with a unit cell of $N$-nuclei and
 electrons moving on a complex potential-energy surface. 

Extraction of `atomic' properties invokes an expansion of,
 say, the potential
felt by an ion at $\vec{R}$ in terms two-body, three-body and higher
interactions. A truncation in finite order and fitting a required
property, e.g., an effective potential, to a large parameter set is used.
If individual atomic properties, e.g., charge state $\bar{Z}$, X-ray
Thomson Scattering (XRTS) profiles etc., are needed, the $N$-atom
cluster has to be decomposed using some
additional model~\cite{Plage-XRTS15,BethkenZbar20} for `decomposing'
the cluster into contributions from individual atomic centers. In
effect, these methods attempt to reconstruct the Neutral-Pseudo-Atom
result from the $N$-atom DFT calculation. Usually  $N$ ranges from 100-500
particles unless crystal symmetry or  some such 
property  can be invoked. 

A kinetic energy functional~\cite{Karas18} may be used to
simplify the DFT computations if some accuracy could be sacrificed, but
`one-atom' properties or pair-potentials are not directly available
from such $N$-ion calculations either.

The explicit dynamical evolution of atomic positions used in large
codes~\cite{ABINIT,VASP} works well for solid-state physics
 applications based on the unit cell, or for the quantum chemistry of
 a finite number of atoms usually at $T=0$. They provide microscopic
 details of `bonding' between atoms not available from the NPA.
 The NPA gives only a thermodynamic average of the system, via the
 thermodynamic one-body densities $n(r),\rho(r)$.
 In constructing the NPA that reduces the $N$ center problem to a single center,
we explicitly identify only one nucleus and then consider the one-body
 distributions $n_s(r), s=e,i$ around it rather than their explicit positions,
 and exploit the spherical symmetry of liquids and plasmas, or
the crystal symmetry of solids, to provide a computationally very efficient
 and yet completely rigorous scheme. The generalization to mixtures
containing many species increases the number of XC-functionals needed, and
may proceed as in Ref.~\cite{eos95}. The NPA theory of mixtures will be
illustrated by some explicit examples.

Use of DFT for the ion distribution avoids the {\it ad hoc} introduction of
multi-atom effects found  in semi-empirical potentials like
the Stillinger-Weber potentials, effective medium models, or in the
Finnis-Sinclair potentials~\cite{Finnis-Sinc09} used in metal physics.
 We have shown that a systematic procedure based on the Ornstein-Zernike
 equation exists for including the three-body and  higher terms into
the ionic XC-functional $F_{ii}^{xc}$. Since ions are classical particles
 in most cases,  the `exchange' content of $F_{ii}^{xc}$ is negligible.

Unlike fitted potentials, the NPA method, being a first principles approach,
can be used for unusual states of matter. 
If the electron subsystem is at a temperature $T_e$, and the ions subsystem is
at a temperature $T_i$, we have one body densities $n_e(r,T_e)$ and
 $\rho(r,T_i)$ defining a two-temperature WDM system.
A quasi-equilibrium exists due to the slowness of the energy relaxation via
electron-ion collisions. Then a quasi-thermodynamic situation exists for time
scales $\tau$ such that $\tau_{ee}\ll \tau \ll \tau_{ii}\ll \tau_{ei}$, where 
$\tau_{ss'}$ is the temperature relaxation time~\cite{cargese94} between
species $s$ and $s'$. Then the NPA approach can be generalized to two-temperature
WDMs and explicit NPA calculations are available~\cite{HarbourDSF18,cargese94}.

Although the potentials in NPA are constructed
in linear response theory, they include the non-linear effects brought in via
the Kohn-Sham atomic calculation. Hence the method is applicable to a wide
range of densities and temperatures. Here we examine simple fluids,
mixtures of fluids, and  complex fluids like  Si, containing transient
covalent bonds, from low $T$ to high $T$. 
Examples where linear-response fails are, for instance, 
liquid-C and liquid transition metals at low $\rho,T$~\cite{Stanek21},
where special procedures are needed, even in standard DFT.

However, in general, it is found that these
NPA  pair-potentials work better than common
multi-center potentials,  and inexpensively recover pair-distribution
functions and other properties in good agreement with the best
available DFT simulations, for well studies systems like 
Al, Be, Li, etc., and for complex fluids like C, Si and Ge.
NPA calculations recover the properties of supercooled liquid Si
at 2.57 g/cm$^3$ just below the melting point of silicon, with
only a fraction of the computation effort needed using VASP-type
$N$-center calculations~\cite{cdwSi20}.

The NPA, based on the two-component DFT of ions and electrons
 enable complex WDM calculations to be reduced to simple one center
 Kohn-Sham calculations that
require nothing more than a laptop since the NPA is computationally very
like an average-atom calculation. But these reach the accuracy of standard
DFT calculations implemented by large codes like ABINIT~\cite{ABINIT} or
VASP~\cite{VASP}, where advanced gradient-corrected meta functionals have
to be used because the $N$-center electron density is very complex and
highly non-uniform.

The NPA approach can impact WDM research as follows.
(i) The method is applicable to metallic solids, liquids or plasmas from
very low $T$ to very high $T$.
(ii) It usually provides reliable one-atom properties like pseudopotentials,
charge states $Z^{j+}$, atomic eigenfunctions, phase shifts, densities of states etc.
 (iii) It provides pair properties  like pair-potentials, 
structure factors and pair-distribution functions (PDFs)
 needed in calculating optical and transport properties of matter. 
(iv) It provides a rigorous, rapid many-body calculation of the ionization balance
and thermodynamics, including composition fractions $x_j$ of mixtures of
ionization states $Z^{j+}$. (v) It has proven useful in
 studying two-temperature non-equilibrium systems and temperature relaxation
rates. (vi) Its rapidity and wide $\rho,T$
 applicability enables easier uncovering of phase transitions etc.,
 as shown recently for liquid silicon at its
melting point~\cite{cdwSi20} and in WDM states,
 or for liquid carbon~\cite{CPP-carb18,DWP-carb90}.
 (vii) It provides a transparent physical model based on DFT,
 without {\it ad hoc} fit parameters. (viii) Its rapidity and
 simplicity enables easy incorporation in large dynamical
 calculations. The limitations of the method are similar to
those of DFT. Additionally, the one-atom DFT implemented in the NPA is
not successful for systems with low free-electron densities at low
$T$. 

In the following we present our theory of pair potentials
as follows. The NPA is introduced as a rigorous DFT concept and then various
simplifications are indicated.  We present a computational procedure for
determining the equilibrium one-body densities $n(r),\rho(r)$,
and the needed XC-functionals.  Then we derive simple linear-response
pseudo-potentials  from the $n(r)$ so obtained. These pseudopotentials
apply to interactions mediated by the continuum electrons,
also called valance electrons, metallic electrons,
or `free electrons'. The atomic cores are described by
the bound states obtained from the NPA calculation. The pseudopotentials and
bound-densities are used to construct bound-bound, bound-free, and free-free 
contributions to the pair potentials. Van der Waals type interactions
occur in the bound-bound interactions. Then we compare our potentials
with SW type multi-center potentials, and also with
effective medium theories.  It is noted that, unlike the short-ranged
potentials of multi-center empirical models, the long-range pair-potentials
of the NPA, with many minima that register with the second, third and higher
neighbour effects in the pair-distribution functions, bring in multi-ion
correlations. An analysis of the potential felt by an ion in a fluid is
used to show how these effects are included via the Ornstein-Zernike equation
and the modified hyper-netted chain equation. Examples are given, including
a comparison with the 40 parameter `glue potential' for aluminum from
effective-medium theory.

\section{Details of the NPA model}
A brief description of the NPA model is given for
convenience and for defining the notation used. Hartree
atomic units ($|e|=\bar{h}=m_e=1$) will be used unless
 otherwise specified.
The systems explicitly studied are  warm-dense fluids and
hence have  spherical symmetry, although the NPA
method  applies equally well to low-temperature
crystalline solids~\cite{Ziman67,Dagens72,Pe-Be}. An
application to surface ablation and hot phonons
has been reported, e.g.,~\cite{HarbEOSPhn17} recently.
However, the early NPA model (e.g., of Ziman)  sought mainly to create 
a computationally convenient short-ranged object named the NPA, so that
inconvenient long-ranged Coulomb fields  of the ionic charges are
neutralized.

The objective of the NPA model used here~\cite{cdw-N-rep19,eos95,Pe-Be} is
to implement a simplification of the full DFT model for electrons
and ions given in Ref~\cite{DWP1} and illustrated by an application to hydrogen.
Its objectives have to be distinguished from early NPA models, or from various
currently-available ion-sphere (IS) or average-atom (AA) 
models~\cite{SternZbar07,Rosznayi08,FausAA10,PironBlenski11,Murillo13,StarretHam14,
YongHou-AA-17} implemented in various  laboratories. Many of them invoke
simplifications which are outside DFT. In fact, the early
formulations used an expansion of the ion-ion interactions in
one-body, two-body, etc., contributions and
 invoked assumptions of non-overlapping electron distributions,
 superposition approximations,
 neglect of higher order ionic and electronic correlations. They also
 invoked the Born-Oppenheimer approximation. The formal  theory
 of the NPA does not need any of these assumptions and includes
higher-order effects via XC-functionals~\cite{DWP1}.
 However, some of these approximations are invoked~\cite{Pe-Be,eos95}
for numerical simplicity.

Our NPA model~\cite{DWP1,Pe-Be,eos95} is
 based on the variational property of the grand potential
 $\Omega([n],[\rho])$ as a functional of  the  one-body densities
  $n(r)$ for electrons, and $\rho(r)$ for ions, even though some
indicated references, e.g., Ref.~\cite{Pe-Be} use a more conventional
 liquid-metal type of presentation.
The Hohenberg-Kohn
theorem applies equally to electrons or ions~\cite{EvansDFT79,DWP1}.
 Only a single nucleus
of the material to be studied is used and taken as the center of the
 coordinate system. The other ions (``field ions'') are replaced by
 their one-body density distribution $\rho(r)$ as DFT asserts that
 the physics is solely given by the one-body distribution; i.e.,
 we do not discard two-body, three-body, and such  information as they
 get included via exchange-correlation (XC)-functionals. That is,
there is no `mean-field' approximation used. Furthermore, 
in the case of ions, the exchange effects are usually negligible
although the ion-ion many-body functional will also be called an
XC-functional for generality.
 
Hence unlike in $N$-center DFT codes like the VASP or ABINIT,
the NPA is a type of one-center DFT using DFT in its full sense.
Consequently, there is no highly inhomogeneous multi-ion potential energy
surface, or a complex  electron distribution
resident on such a multi-center potential energy surface.
Hence a local-density approximation (LDA)
to the electron XC-functional can be sufficient and no gradient
expansions are needed. In our calculations
we have used the finite-$T$ electron XC-functional by Perrot
and Dharma-wardana~\cite{PDWXC} in LDA, while most large-scale codes
implement  $T=0$ XC-functionals like the PBE
functional~\cite{PBE96}. These include generalized gradient
corrections, and other  advanced but computationally very demanding
functionals, e.g., the SCAN functional~\cite{SCAN13}. It is 
seen that our NPA-LDA calculations are in close agreement with $N$-atom
DFT calculations using the PBE or SCAN functionals even for complex
transiently bonded liquids like molten Si~\cite{cdwSi20} or
liquid carbon~\cite{CPP-carb18}, with the VASP results disconcertingly
dependent on the choice of the XC-functional~\cite{Remsing17}. 
Our finite-$T$ functional
used is in good agreement with quantum Monte-Carlo XC-data~\cite{cdw-N-rep19}
in the density and temperature regimes of interest.
 
The artifice of using a nucleus at the origin converts the one-body ion
 density $\rho(r)$ and the electron density $n(r)$ into effective two body
 densities in the sense that
\begin{equation}
\rho(r)=\bar{\rho}g_{ii}(r),\; n(r)=\bar{n}g_{ei}(r).
\end{equation}
The ion at the origin need not be at rest. However, most ions are heavy enough that
the Born-Oppenheimer approximation is valid.
Here $\bar{\rho},\bar{n}$ are the average ion density and the average
 free electron density respectively, and prevail far away from the
central nucleus. Any bound electrons are 
 assumed to be firmly associated with each ionic  nucleus
and contained in their ``ion cores'' of radius $r_c$ such that
\begin{equation}
\label{core.eqn}
 r_c<r_{\rm ws},\; r_{ws}=\left[3/4\pi\bar{\rho}\right]^{1/3}. 
\end{equation}
Here $r_{\rm ws}$ is the Wigner-Seitz radius of the ions. The corresponding
electron sphere radius, based on $\bar{n}$ is denoted by $r_s$. 
In some cases, e.g., for some transition metals, and for continuum
 resonances etc., this condition for a compact core may not be met,
 and additional  steps are needed. We assume that $r_c<r_{ws}$
 unless stated otherwise. The DFT variational equations used here are:
\begin{eqnarray}
\label{KS-basic.eqn}
\frac{\delta \Omega[n,\rho]}{\delta n}&=&0 \\
\label{KS-ion.eqn}
\frac{\delta \Omega[n,\rho]}{\delta \rho}&=&0.
\end{eqnarray}
These directly lead to two coupled KS equations where the
unknown quantities are the XC-functional for the electrons,  the
 ion XC-functional, and the electron-ion XC functional~\cite{ilciacco93}.
 If the Born-Oppenheimer approximation is imposed, the electron-ion
 XC-functional may also be neglected. Approximations arise in
 modeling these functionals and
in  simplifying the decoupling of the two KS  equations~\cite{eos95,CPP15}.
 The first equation gives the usual
 Kohn-Sham equation for electrons moving in the external potential of
 the ions. This is the only DFT equation used in the $N$-center DFT-MD
 method implemented in standard codes like the VASP, where the ions define
a periodic solid whose structure is varied via $N$-atom classical
 MD, followed by a Kohn-Sham solution at
 each step. 

In the NPA the one-body ion density $\rho(r)$ rather than
an $N$-center ion density is used. It was shown in ~\cite{DWP1}
that the DFT  equation for the one-body ion density can be identified
 as a Boltzmann-like  distribution of field ions around the
 central ion, distributed according to the `potential of mean force'
 well known in the theory of fluids. Hence the 
ion-ion correlation functional $F_{ii}^{xc}$ was identified to be
exactly given by the sum of
 hyper-netted-chain (HNC) diagrams plus the bridge diagrams.
The bridge diagrams have to be approximated
from a hard-sphere fluid~\cite{LFA83}, or evaluated using MD.

The mean electron density  $\bar{n}$ can also be specified as the number of
free electrons per ion, viz., the mean ionization state $Z$. It is also
denoted as $\bar{Z}$ since a plasma containing a mixture of ionization
states $Z^{j+}$ with composition fractions $x_j$ will appear as a single
average ion of charge $\bar{Z}=\sum_j x_jZ^{j+}$. This average description
can be sharpened by constructing individual NPA models for each
integer ionization and using them in the thermodynamic average~\cite{eos95}.
Hence the NPA approach yields the composition fractions $x_j$ in the plasma,
although the ABINIT and VASP calculations {\it cannot} yield such
 information directly. Unfortunately,  the
 partitioning  of the $N$-atom density from $N$-atom results into individual atomic
 contributions is not unique and a variety of schemes
 exist~\cite{Li-Parr86,Francisco06,BethkenZbar20}.

Although the material density $\bar{\rho}$ is specified, 
the mean  free electron density $\bar{n}$ is initially
unknown at any given temperature, as it depends
 on the  ionization balance which is controlled by the free energy
 minimization implied by the DFT Eq.~\ref{KS-basic.eqn}. 
Hence a trial value for $\bar{n}$, i.e, equivalently, a trial value for
  $\bar{Z}$, is assumed in the NPA and the thermodynamically consistent $\rho(r)$ is
determined. This is repeated till the target mean ion  density $\bar{\rho}$
 is obtained. This also implies that  NPA implements a
calculation of  the ionization balance in the system. In effect, a
full many-body extension of the Saha equation via its DFT free-energy
minimum property is implemented in the NPA model.

The mean number of electrons per ion, viz., $\bar{Z}$, is often replaced by
$Z$ in this paper when no confusion arises. It is an experimentally  measurable
quantity. It is routinely measured using Langmuir probes~\cite{Smy76}
in atmospheric and gas discharge plasmas, or from optical and other
measurements of appropriate properties, e.g., the conductivity and the
XRTS  profile~\cite{GlenzerRedmer2009} for general WDM systems. The
argument that `there is no quantum mechanical operator' corresponding
to $\bar{Z}$, and hence it is not an observable, is incorrect. The system
studied is not a pure quantum system but a  system
attached to a classical heat bath.  Here $T,\mu, \bar{Z}$
are Lagrange multipliers associated with the conservation of energy, particle
number, and charge conservation respectively. They do not correspond to
simple quantum operators whose mean values are $T,\mu$ or $\bar{Z}$, although
complicated operator constructions can be given. 

\subsection{Computational simplifications}
The Kohn-Sham equation has to be solved for a single
 electron moving in the field of the central ion surrounded by $n(r),\rho(r)$.
 The ion distribution
 $\rho(r)=\bar{\rho}g(r)$ associated with it is also modified at each iteration
 with a corresponding  modification of the  trial $\bar{Z}$ until the
 target density $\rho$ is obtained. However, it was noticed  very
 early~\cite{Pe-Be, eos95} that the Kohn-Sham
 solution was quite insensitive to the details of the $g(r)$ in most
cases; an example of an exception  being  low-density liquid carbon at low $T$.
 Hence a
 simplification was possible~\cite{DWP1,Pe-Be,eos95}. The simplification was
 to replace the trial
 $g(r)$ at the trial $\bar{Z}$ by a cavity-like distribution:
\begin{eqnarray}
\label{cavity-gr.eq}
g_{cav}(r)&=&0, r\le r_{\rm ws}, \;  g_{cav}(r)=1, \; r>r_{\rm ws}\\
\rho_{cav}(r)&=&\bar{\rho}g_{cav}(r), \\
V_{cav}(k)&=&V_k\bar{Z}(3/K)^3\left[\sin(K)-K\cos(K)\right]\\
V_k&=&4\pi/k^2, \; K=kr_{\rm ws}.
\end{eqnarray}
Here the $r_{\rm ws}$ is the trial value of the ion Wigner-Seitz radius,
 based on the trial $\bar{n}$. Convergence of the equations using the
cavity approximation instead of the full $g(r)$ is very rapid, since
 adjusting the $g_{cav}(r)$ at each
 iteration requires only adjusting the trial $r_{\rm ws}$ to achieve
 self-consistency. The self-consistency in the ion distribution is
 rigorously controlled by the Friedel sum rule for the phase shifts of
the KS-electrons~\cite{DWP1}. This ensures that $\bar{\rho}=\bar{n}/Z$.
Thus, a valuable result of the calculation is the self-consistent
 value of the mean ionization state $Z$,
 which is both an atomic quantity and a thermodynamic quantity. Here too
we note that a  sophisticated  quantum many-body version  of a Saha equation
is  also solved automatically within this DFT calculation in
 determining $\bar{Z}$.
The pseudopotentials, pair-potentials, the equation of state (EOS) etc.,
depend directly on the accuracy of $\bar{Z}$. In contrast, many average-atom
models use various prescriptions based on ion-sphere models (valid at high-$T$)
to determine $\bar{Z}$~\cite{SternZbar07,Murillo13}.

In early discussions of the neutral pseudoatom
as applied to solids~\cite{Dagens72,Ziman67}, no attempt was made to introduce a
DFT of the ions, and an analytically convenient potential neutralizing the
long-range Coulomb interaction of the central ion
was added to produce a computationally
 convenient object with a short-ranged potential. In such studies, and in
MD simulations of Coulomb potentials, a neutralizing  Gaussian charge
distribution is sometimes used. The spherical cavity model used here
is a good approximation to a typical ion-ion $g(r)$ for $r< r_{ws}$.  

Here  we note  several simplification used in implementing
the NPA. Given that the electron distribution $n(r)$ obtained self 
consistently can be  written as a core-electron (i.e., bound electron) term
and a free-electron term when the core electrons are compactly
inside the ion Wigner-Seitz sphere, we have:
\begin{eqnarray}
\label{pileup.eqn}
n(r)&=&n_c(r)+n_f(r), \; \Delta n_f(r)=n_f(r)-\bar{n}\\
a_f(r)&=&\Delta n_f(r).
\end{eqnarray} 
The core-electron density (made up of ``bound electrons'') is denoted by
$n_c(r)$. The free electron density $n_f(r)$ is the response of an
 electron fluid~(see sec.~\ref{uef.sec}) to the central ion
  and to an ion-density deficit (a cavity) near the central ion
 that mimics  $\rho(r)$. It contributes
 to the potential acting on the electrons. The response of a uniform electron
 fluid (i.e., without the cavity) to the central ion can be obtained by
  subtracting out the effect of the cavity using the known static
 interacting linear response function  $\chi(k,\bar{n},T)$ of the
 electron fluid. That is, from  now on we take it that the charge
 density $n_f(r)$ and the density pile up  $\Delta n_f(r)$ are both
 corrected  for the presence of  the cavity.
 However, we use  the same  symbols as before when no confusion arises.
 Furthermore, in dealing with atoms `a', `b' etc, we refer to the
 `density pileup'  or {\it displaced} free
electron density $\Delta n_f(r)$ as $ a_f(r)$, with its
 Fourier transform denoted by $ a_f(k)$. 

\begin{figure}[t]
\begin{center}
\includegraphics[width=\columnwidth]{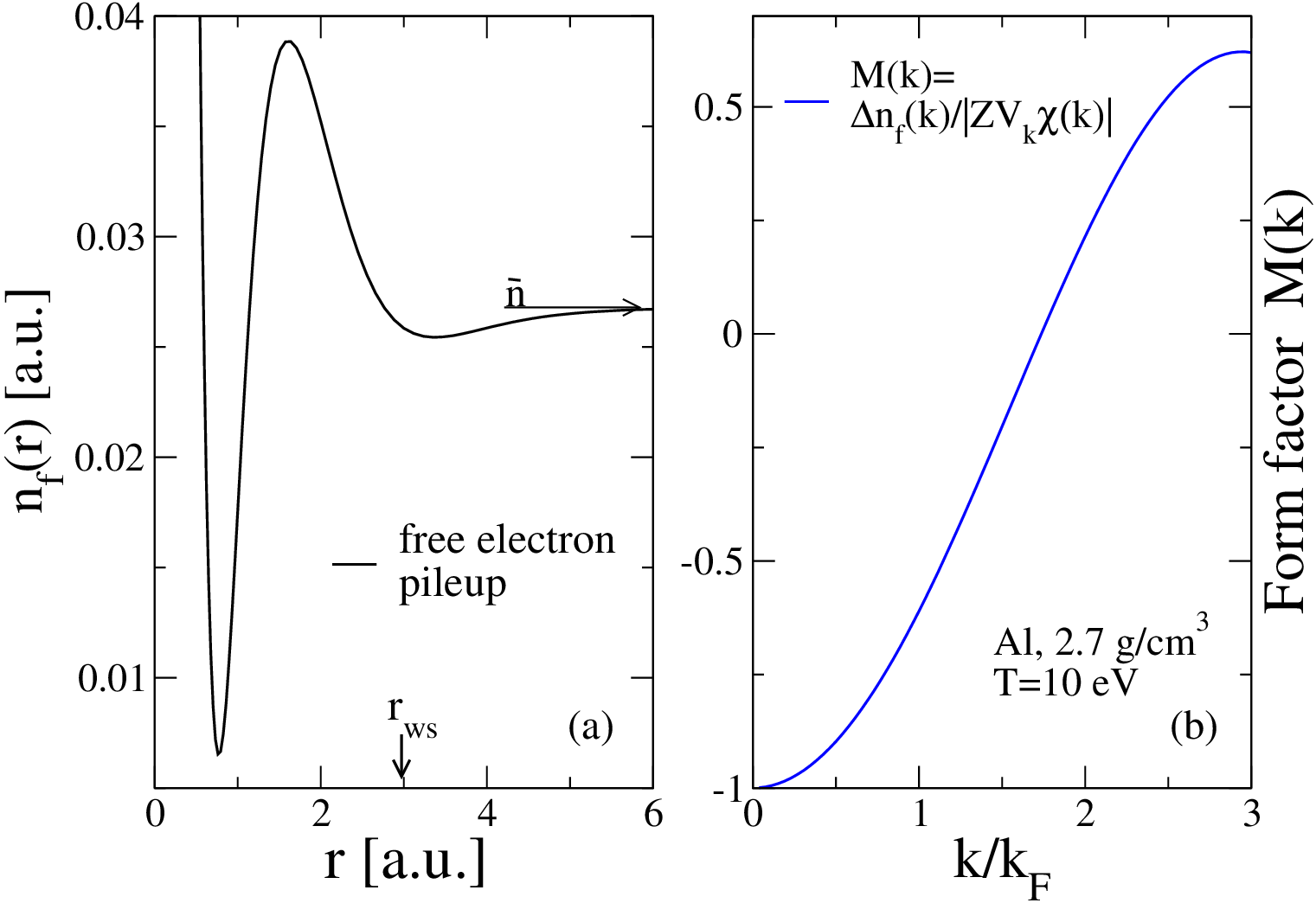}
\caption{
\label{formfac.fig}
(Online color)  Panel (a). The free electron density $n_f(r)$ around an
Al nucleus in its electron fluid of average density
$\bar{n}$ corresponding to $\bar{Z}=3$, $\rho=2.7$g/cm$^3$, at 10 eV. The
bound electron core is inside $r_{ws}$, and $n_f(r)$
develops oscillations as the free-electron states are orthogonal to
core states. Panel (b). The $k$- dependent pseudopotential
form factor $M(k)$, calculated from the electron pileup $\Delta n_f(k)$
and the electron response function $\chi(k,T)$. Here $V_k$ is
the Coulomb potential $4\pi/k^2$. The large $k/k_F > 2$ behaviour
is unimportant unless $T> E_F$. Here $k_F$ is the Fermi wave vector.
}
\end{center}
\end{figure}

In Fig.~\ref{formfac.fig}(a) we display the calculated $n_f(r)$
for an Aluminum ion in a hot plasma.
Unlike in ABINIT or the VASP where core electrons are subsumed
in a pseudopotential, the NPA uses
all-electron atomic calculations and provides the true core-electron
density as well.

A key difference between many average-atom models and the NPA is that
 the free electrons are not confined to the Wigner-Seitz sphere, but
 move in all of space represented by a very large `correlation
 sphere' of radius $R_c$. This may be ten to twenty times the Wigner-Seitz
 radius of the central ion~\cite{eos95,Murillo13}. This ensures that
the small-$k$ limit of the structure factor $S(k)$
is accurately obtained.

\section{Pseudopotentials and Pair-potentials from the NPA model.}
\label{pots-npa.sec}
As the NPA reduces the multi-center WDM to a single ion and its environment
described by $n(r),\rho(r)$, pair potentials must be constructed from
the appropriate one body distributions, viz, the core-electron
density $n_c(r)$, and the `free' electron
density $n_f(r)$.

 The pseudopotential $U_{ei}(r)$, or its Fourier transformed
form $U_{ei}(k)$ enables a rigorous and useful separation of the
contributions of bound and free  charge densities to the pair potential
 which is a two-center property. We constructed it from one-center
results to avoid two-center computations. Thus the pseudopotential is
constructed to be a linear response property where possible. In fact,
 unlike in crystals, the spherical symmetry of fluids
allows one to use simple local pseudopotentials, without having to
deal with angular-momentum dependent (i.e., nonlocal) pseudopotentials.
So these potentials use only an $s$-wave component and differ from
those used in the $N$-center DFT codes. They are also
constructed to be  weak potentials,
allowing the use of second-order  perturbation theory with them. 

Unlike non-linear pseudopotentials used in
ABINIT or VASP, these pseudopotentials are $\rho,T$  dependent, 
and constructed {\it in situ} during the calculation. 
The range of validity of such linearized pseudopotentials  has
been discussed elsewhere~\cite{cdw-Utah12}.

A rigorous discussion of pair-potentials $U_{\rm ab}$ between two atoms
of species `a', and `b', including the effect of core electrons is given in
the Appendix B of Ref.~\cite{eos95}. Core-core interactions are important
for atoms including and beyond argon, viz., Na, K, Au,  etc., with
large cores and zero or low $\bar{Z}$. In contrast, core-core interactions
in high $Z$ ions like Al$^{3+}$, Si$^{4+}$ at  normal density are very
small in comparison
 to the interactions via free electrons.
 
We illustrate the calculation of core-core interactions
by a discussion of the weakly ionized argon WDM system.
The core electron distributions $n_c(r)$ of the atom `a' is denoted by $a_c$,
 while $a_f$ denotes the {\it displaced} free electron distribution, $\delta n_f(r)$,
for the atomic center of type `a' (and similarly for `b').
The total pair-interaction $V_{ab}$
 is of the form:
\begin{eqnarray}
\label{ppot.eqn}
V_{\rm ab}&=&V(a_c,b_c)+ V(a_c,b_f)+V(a_f,b_c)+ V(a_f,b_f)
\nonumber  \\
         &=&V^{cc}(r)+V^{cf}(r)+V^{ff}(r).
\end{eqnarray}
These terms will be discussed separately.

\subsection{The interaction mediated by free electrons}
The last term of Eq.~\ref{ppot.eqn}, viz., $V^{ff}(r)$
 is the familiar ion-ion interaction
mediated by metallic electrons. This is adequately evaluated in
 second-order perturbation theory when the interactions
are weakened by electron screening.
The linearized  electron-ion
pseudopotential described below can be used for WDM systems and
 liquid metals unless
 the free electron density and the temperature ($T/E_F$) are  low,
 when linear-response methods  become invalid. In fact, when the
 pair-potential develops a negative region significantly larger
 than the thermal energy, permanent chemical bonds are formed;
 then the liner-response methods used here cannot be used. 

 The electron-ion pseudopotential $U_{\rm a}^{ei}(k)$ of the
ion `a' is evaluated in $k$-space via the displaced free-electron density
$a_f(k)$, and the electron response function $\chi(k,T)$.
The pseudopotential is:
\begin{eqnarray}
\label{pseudo.eq}
U_{\rm a}^{ei}(k)&=&a_f(k)/\chi(k,T)\\
                 &=&-Z_aV_kM(k),\;V_k=4\pi/k^2.
\end{eqnarray}
Here $M(k)$ is the form factor of the
 pseudopotential (see Fig.~\ref{formfac.fig}). The NPA
calculation via Eq.~\ref{pseudo.eq} automatically yields
the pseudopotential inclusive of a form factor. This may be
fitted  to a Heine-Abarankov form which fits a core
radius $r_c $ and a constant core potential $V_c $. 
Alternatively, a more complex parameterized form 
may be needed, as in Ref.~\cite{cdw-Utah12}. 
 However, using the numerical tabulation of
$U_{\rm a}^{ei}(k)$ is more accurate and avoids the fitting
step. 

The fully interacting linear response function $\chi(k,T)$
of the uniform electron fluid at the temperature $T$ is discussed
in sec.~\ref{uef.sec}. So,  for identical ions `a', we have
the form:
\begin{eqnarray}
\label{aapot-f.eqn}
V_{\rm aa}^{ff}(k)&=&Z_a^2 V_k+|U_{\rm a}^{ei}(k)|^2 \chi(k,T)\\
                  &=& Z_a^2 V_k+|a_f(k)|^2 / \chi(k,T). 
\end{eqnarray}
In the case of two different atomic species, we readily have:
\begin{eqnarray}
\label{ppot-f.eqn}
V_{\rm ab}^{ff}(k)&=&Z_aZ_bV_k +\\
         & & |U_{\rm a}(k)U_{\rm b}(k)|\chi(k,T).
\end{eqnarray}
In the NPA theory for mixtures, $Z_a$ and $Z_b$ are integers, while
in the simple average-atom form of the NPA, the mean value
 $\bar{Z_s}=\Sigma_s x_sZ_s$ is used to represent the mixture 
with composition fractions $x_s$ via a single calculation. In
a mixture calculation, separate NPA calculations for each
component are needed.

The simplest pseudopotential, valid  for point ions, $r_c=0$,  uses a
form factor of unity.  The long-wavelength form ($k\to 0$)
 of the response function applies far away ($r\gg r_{ws}$)
from an ion in the uniform-density region of a plasma. Then it depends only
 on the screening wavevector $k_s(T)$. Given such approximations,
 the electron response function $\chi(k,T)$ and the pair-potential
 reduce to very simple forms.
\begin{eqnarray} 
\label{yukawa.eqn}
\chi(k)&=&k_s^2/4\pi,\; V_{\rm ab}^{ff}= 4\pi Z_aZ_b/(k^2+k_s^2)\\ 
\label{ks.eqn}
k^2_s &=&k^2_{se}(T)=\frac{4}{\pi T}\int_0^\infty k^2dk n(k)\{1-n(k)\}\\
  n(k)&=&1/\left[1+\exp\{(k^2/2-\mu(T))/T\}\right] \\
V_{\rm ab}^{ff}(r)&=&(Z_aZ_b/r)\exp(-k_sr).
\end{eqnarray}
Here  $k_s=k_{se}$ is an electron screening wave vector in the fluid at the
temperature $T$ and at the electron chemical potential $\mu(T)$.
This reduces to the Thomas-Fermi value at $T\to 0$, and to the
Debye-H\"ukel value as $T\to\infty$. 
However, most scattering processes for $T/E_F\le 1$ occur with a momentum transfer
of 2$k_F$  within a thermal window of the Fermi energy.
Hence the use of the $k\to 0$ approximation is limited
to high $T$. 

The real space form of the long-wavelength screening formula
is well known as the ``Yukawa  potential'',
 $Z_aZ_b\exp(-k_s r)/r$. It has
been used  in particle physics, physical chemistry
and in plasma applications 
because of its 
simplicity and validity at weak coupling
~\cite{Rogers70,StanMur16,YukawaEdwards17,SunDai-Yuk17,Stanek21}.

\begin{figure}[t]
\begin{center}
\includegraphics[width=\columnwidth]{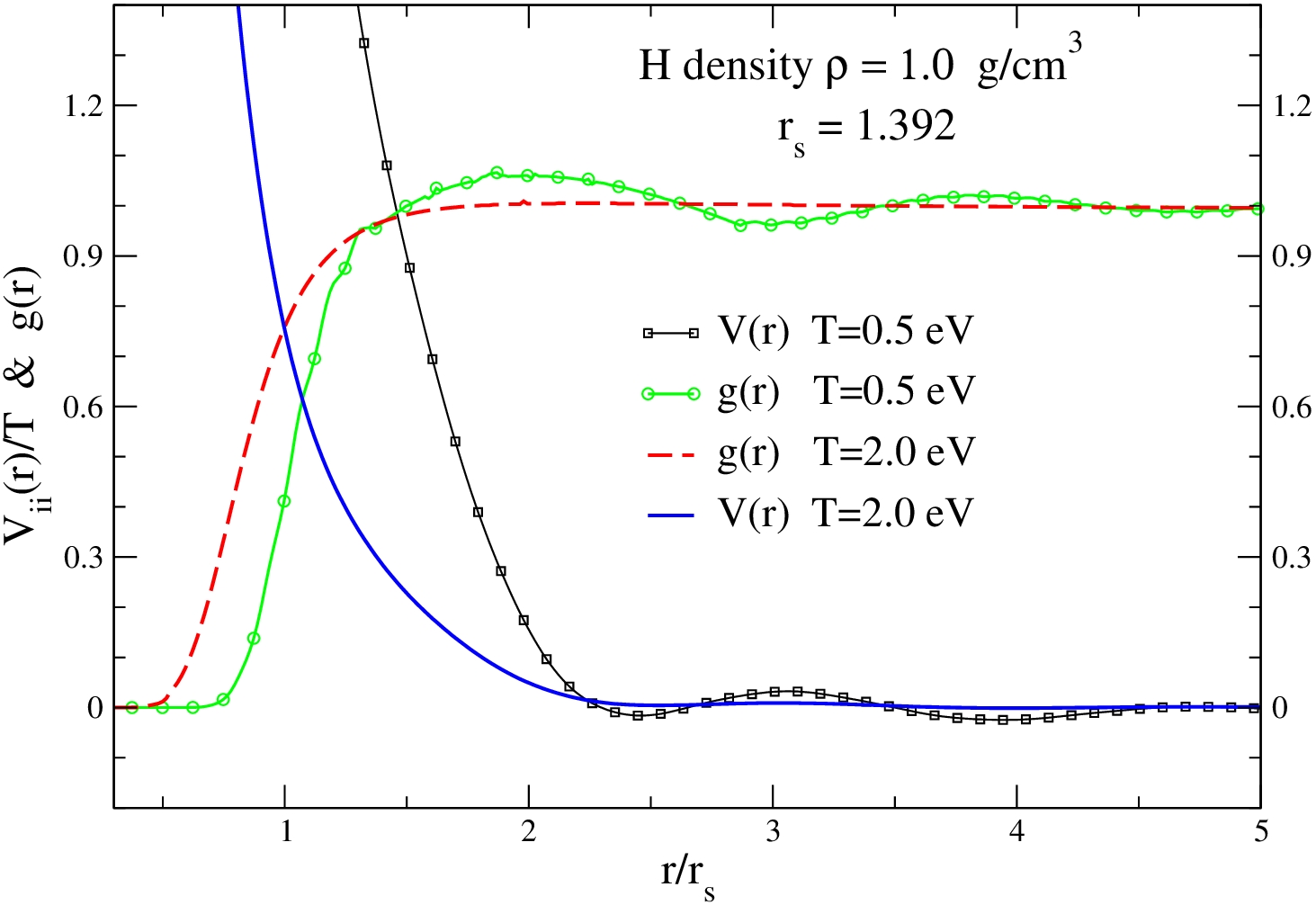}
\caption{
\label{Hydpots.fig}
(Online color)The proton-proton pair potentials and PDFs
at 0.5 eV and 2 eV in a Hydrogen plasma at $r_s$=1.391,
 with $\rho$=1.0 g/cm$^3$, at $T$ significantly
higher than a possible phase transition and contains
protons and electrons.The pair-potentials
 contain Friedel oscillations
whose minima line up with the peaks in the $g(r)$.
}
\end{center}
\end{figure}

Quantum statistical potentials have also been considered for
finite-$T$ electronic systems since the 1960s~\cite{DunnBr67,Filinov04}, and more
recently~\cite{JonesMur07}, where the latter authors concluded that such
methods treat many-body effects inadequately.

The point-ion model is unsatisfactory even for protons for $T<E_F$ as
discussed below.
The proton-proton pair-potential is a most demanding case for the NPA
method because the charge pile up around the proton is highly non-linear,
and the methods used here become inapplicable without further modification
when very few free electrons are present, while partial ionizations strictly
require the use of a mixture of ions of different ionizations. Forms like
 H$^-$, H, H$^+$ as well as quasi-molecular transient forms like H$_2^+$, may occur
in H plasmas near  a phase transition. The fully ionized case (i.e., a proton) has
 no bound core, and yet the point-ion model fails since a
form factor associated with the non-linear charge pile up is needed. We
find that the form factor from the NPA  works with adequate accuracy, and even
picks up the effects of transient H$_2^+$ in the medium when relevant. We
give an illustrative example  below, using fully ionized
hydrogen. Similar transients in hydrogen have been
 noted by Norman {\it et al}~\cite{Norman17H}.

\begin{figure}[t]
\begin{center}
\includegraphics[width=\columnwidth]{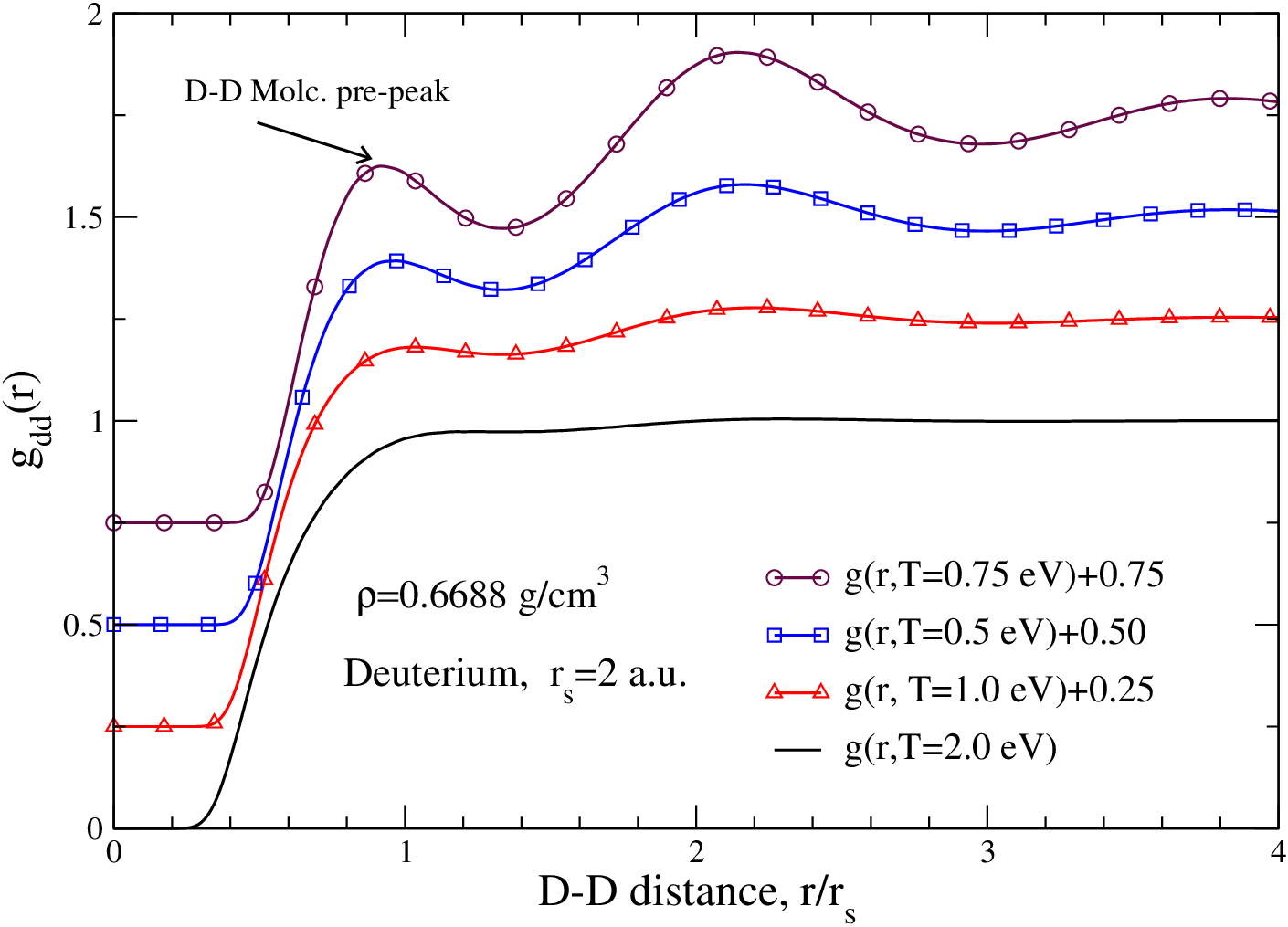}
\caption{
\label{gDD.fig}
(Online color)The deuteron-deuteron pair distribution
functions display the onset of transient D-D molecular forms
 (e.g., D$_2^+$)
as the temperature is lowered from 2 eV to 0.3 eV, for a
deuterium plasma at $r_s=2$. The D$^+$ are fully ionized
and only transient molecular forms exist, while there are no
atomic bound states.
}
\end{center}
\end{figure}

Fig.~\ref{Hydpots.fig} displays the  the proton-proton pair potential and PDFs at
0.5 eV and 2 eV for a hydrogen plasma at $r_s \simeq 1.392$, $\rho=1.0$
g/cm$^3$. The plasma is fully ionized, although
close to  a first-order transition from a molecular
 liquid to a conducting atomic fluid, believed to occur near
$T\sim$ 0.13 eV.  Nuclear quantum effects are also important near the
phase transition.
 How they may be included in NPA models will be discussed elsewhere,
 following previous work based on the classical-map approach
 to quantum effects~\cite{cdw-N-rep19}. Here we neglect
them, working at a higher temperature of 0.5 eV and 2 eV.
In Fig.~\ref{gDD.fig} we display the D-D PDFs at a somewhat
lower density where the tendency to form transient bonding is
higher.

The broad hump in the $g(r)$ in Fig.~\ref{Hydpots.fig} from  1.5$<r/r_s<2.5$ 
corresponds to the range of bond lengths associated with the presence of
transient H$_2^+$ states as well as with H$^+$ packing effects.
The peak due to purely packing effects occurs near $r/r_{\rm ws}\sim 1.6$. 
The nominal H$_2^+$ bond length of $\simeq 2$ a.u. in a vacuum at $T=0$
is extended in a plasma due to screening and temperature effects.
The bond becomes transient, with a broader range of lengths.

It should be noted that the NPA pair-potentials
are very long-ranged, encompassing some tens of  Wigner-Seitz radii.
The aluminum pair-potential near its melting point requires
using a pair-potential extending over at least a dozen atomic shells
if the phonon spectra are to be accurately recovered~\cite{HarbEOSPhn17}.
This should be compared to those used in the pair-potential
part of empirical models like the SW potential which extends to only
about one Wigner-Seitz radius. The belief popular in the empirical
modeling community that pair-potentials cannot reproduce tetrahedral
structures without ``three-body forces'' is certainly true if the
pair-potential  does not even extent to the next-nearest
neighbour (N-N-N) position! 

 The
minima in the Friedel oscillations of the NPA potentials
correspond approximately to shells of atoms where
the third, fourth and higher neighbours are positioned, as seen 
by a comparison with the PDFs generated from these potentials.
The PDFs are generated using the NPA potentials in classical
MD, or in a hyper-netted-chain equation inclusive of a bridge term.
Simulation methods for long-range potentials, viz,,  Ewald
constructions, use of analytic tails etc., are well known
and pose no difficulties, even for extracting dynamic
structure data using NPA potentials~\cite{Nadin88,HarbourDSF18}. 

A potential with wider applicability to uniform fluids than the
Yukawa model is obtained
using the random phase approximation (RPA) to the response function. This
retains the all important Friedel oscillations in the pair potential for
$T/E_F < 1$.  
 At high temperatures this reduces to the Vlasov approximation and then to
the Yukawa form. 
The  pair-potential, the detailed computations of the pair distribution
functions (PDFs), free energies and  EOS properties
of a plasma of point charges interacting via
the  RPA screened Coulomb potential is an
important reference point. These have been computed
 as a function of the plasma parameter $\Gamma=Z/(r_{ws}T)$
 by Perrot~\cite{Perrot91}. However, the
 assumption of negligible core size (point-ion approximation)
is not adequate for most ions except at high $T$.
Neither the RPA form, nor the Yukawa form, is designed to satisfy
the compressibility sum rule.

A  complete treatment of
ion-ion interactions needs core-core interactions as well as the inclusion
of a form factor in the electron-ion pseudopotential $U_{\rm a}^{ei}(k)$.
We use Eq.~\ref{pseudo.eq}, which is an accurate yet simple
pseudopotential directly available from any average-atom
calculation. We have successfully used the NPA $U^{ei}(k)$ 
in many diverse materials and WDM systems ranging from solid Be, Na, Al, 
to plasma states of Be, Li, C, Ga, Al, or Si even as a supercooled liquid. 

The pseudopotential resulting from Eq.~\ref{pseudo.eq} is a simple
local potential ($s$-wave potential). If the density displacement
 $a_f(k)$ is expanded in terms of the contributions from different
 angular momentum states $l$ of the
Kohn-Sham eigenstates, then a non-local ($l$-dependent) pseudopotential
can be constructed. Such non-local forms are
needed in solid-state applications where spherical symmetry is lacking.
However, the simple $s$-wave
form given in Eq.~\ref{pseudo.eq} seems to work well for fluids,
their EOS and static transport coefficients.

\begin{figure}[t]
\begin{center}
\includegraphics[width=\columnwidth]{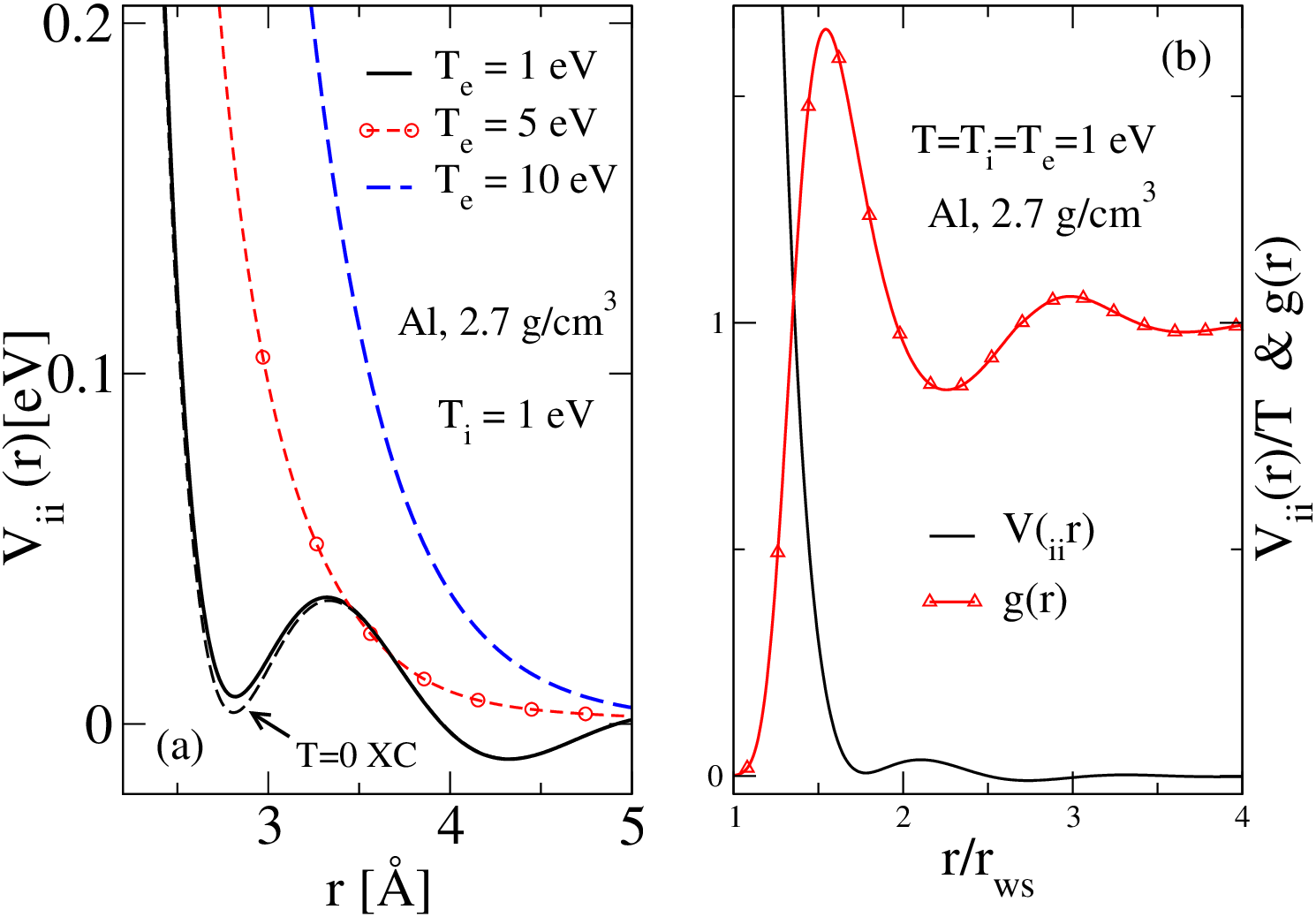}
\caption{
\label{Al-Vr.fig}
(Online color) (a)The Al-Al pair potentials at normal density
from the NPA for WDMs in thermal equilibrium at $T=1$ eV, and 
two WDM states  with $T_e\ne T_i$. The Fermi energy $E_F\sim 12$ eV, and
the electron degeneracy changes. If the
electron XC is treating using the $T=0$ instead of $T=1$ eV,
 there is a slight error in the depth of the
first energy minimum. (b) The Al pair-potential  at 1 eV
 (black line) and its g(r) are shown as a red line
with triangles. The energy is scaled by $T$, and the position
by $r_{\rm ws}$, unlike in the left panel.
The first maximum in the g(r) is at a {\it positive}
energy and not at the lowest minimum in the pair potential.
It is the preferred first-neighbour position  because
the  the XC-energy favours a denser  electron fluid.
The second and third neighbours, corresponding to the secondary peaks
in the PDF are nearly at the Friedel minima.
}
\end{center}
\end{figure}

Aluminum is regarded as a 'difficult' material by those who develop
effective medium potentials for it. In Fig.~\ref{Al-Vr.fig} we display
Al-Al pair potentials for several temperatures and compressions,
including for a case where the
ions are held at 1 eV, while the electrons are at 1 eV, or at 5 eV and 10 eV.
The potentials show how the Friedel oscillations disappear at high electron
temperatures. These minima, which determine the location of N-N,
N-N-N  positions bring in the multi-center features that
empirical models like the SW potential put in ``by hand''.
 However, Fig.~\ref{Al-Vr.fig}
shows that the peaks and troughs in the $g(r)$ are not entirely determined by
the pair potential as the `volume energy' of the electron fluid plays a
part, and largely determines the position of the first peak in high-$\bar{Z}$
 systems like
liquid aluminum or liquid carbon (see discussion in Ref.~\cite{DWP-carb90}).

\begin{figure}[t]
\begin{center}
\includegraphics[width=\columnwidth]{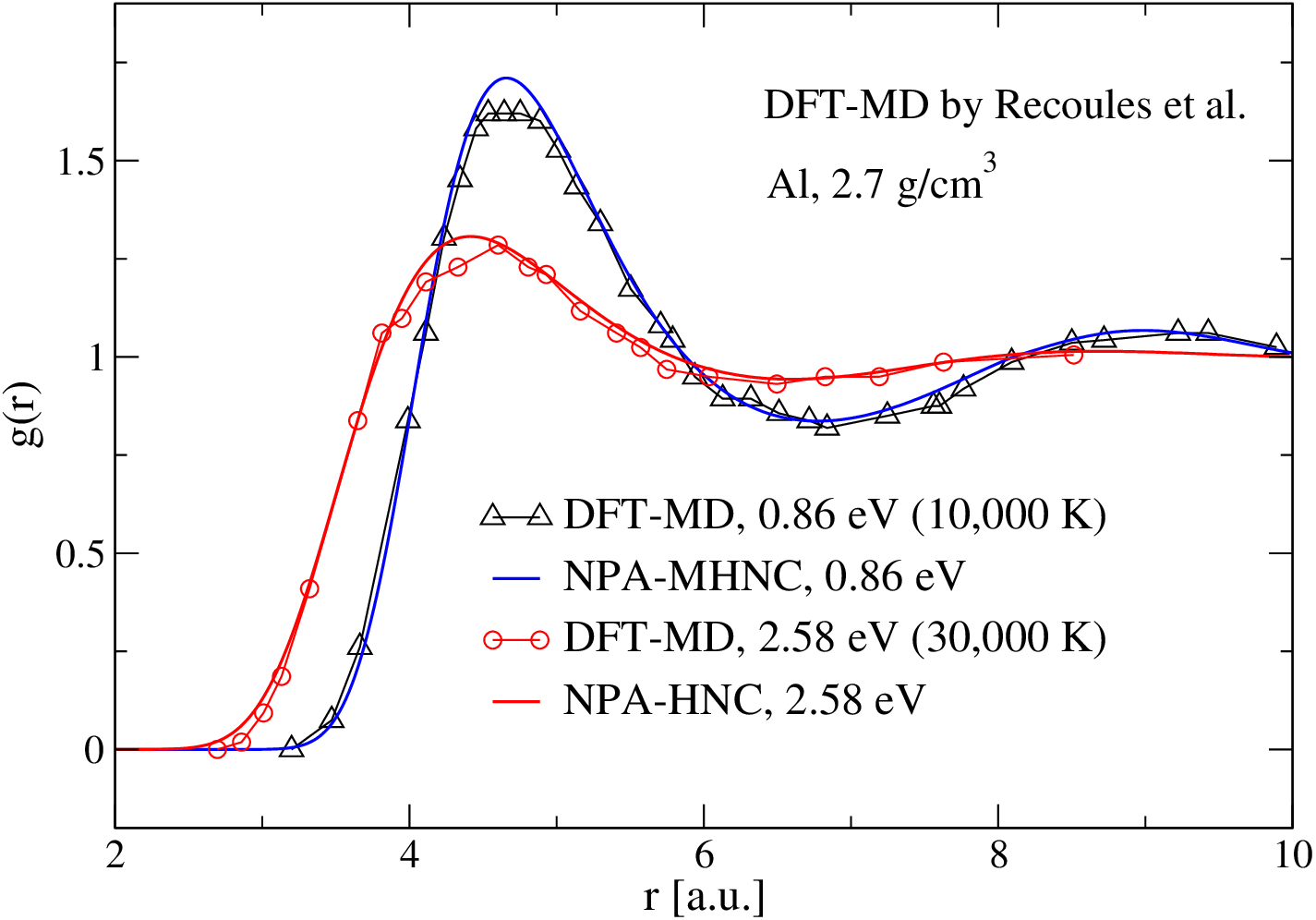}
\caption{
\label{Algr.fig}
(Online color)The NPA $g(r)$ for aluminum at two temperatures
are compared with the DFT-MD simulations of Recoules {\it et al}~\cite{recoules15}.
The simulations use 64 atoms and the PBE functional. Better agreement near
the first peak is obtained from simulations with more atoms.
}
\end{center}
\end{figure}

The PDFs $g(r)$ can be obtained from the pair-potential either using
classical MD, or using the modified HNC equation (MHNC). The $g(r)$ shown in
Fig.~\ref{Al-Vr.fig}(b) has been obtained using the MHNC
 and a Bridge term based on
a hard sphere liquid with a packing fraction of 0.2996. It is also of
 interest to see how well these NPA PDFs of one-center DFT agree with
 direct simulations using an $N$-center DFT procedure like that of the
VASP or ABINIT. In Fig.~\ref{Algr.fig} we give comparisons of NPA for
two PDFs  with those of Recoules {\it et al}.~\cite{recoules15}, with $N$=64
atoms in the simulation cell. The PDFs
obtained from DFT-MD simulations are sensitive to the type of electron
XC-functional used, and a higher $N$ is needed to get better statistics.
However, the NPA $g(r)$ and the DFT-MD can be considered to be in good 
agreement.  

\subsection{Contributions to the potential from the core-electron density}
\label{core.sec} 
Equation ~\ref{ppot.eqn} contains $V^{cc}$ and $V^{cf}$ where the core-electron
density contributes to the pair-potential. In the following we show that
the neglect of core-core interactions can lead to very serious differences, 
e.g., specially in weakly ionized plasmas containing neutral species.

The term  $V^{cc}(r)$  contains $V(a_c,b_c)$ which is the
interaction between the density distributions of the bound electron core
in the atom `a' with the core electrons in `b'. These are not the
 unperturbed NPA
 densities $n_c(r)$ defined in Eq.~\ref{pileup.eqn}, but the densities
 that result from the perturbation of each density by the other,
 in the plasma environment.
 The unperturbed core density of the
 ion `a' is denoted by $n^{0}_c(\vec{r}_a), \vec{r}_a=\vec{r}-\vec{R}_a$
 and similarly for the ion `b', with $R =|\vec{R}_a-\vec{R}_b|$
 defining the  scalar separation of the two atoms. The
 contribution to the pair-potential is:
\begin{eqnarray}
\label{cc-pot.eqn}
V^{cc}(R)&=&Z_aZ_b/R+
\int \frac{dr^3 dr'^3 n^{0}_c(\vec{r}_a)n^{0}_c(\vec{r'}_b)}
{|\vec{r}-\vec{r}'|}+\nonumber\\
 & &\Delta_{ES}(R)+\Delta_{xc}(R).
\end{eqnarray}
The potential contains an electrostatic correction
 $\Delta_{ES}(R)$ as the core densities are modified by the interaction.
Instead of doing a two-center calculation for this term, second-order
perturbation theory based on the polarisability of the
core density is sufficient specially when there are free-electron 
screening effects, as in plasmas or metals.
An electron XC-term $\Delta_{xc}(R)$ is associated with the
 density modification.
This includes a correction to the kinetic energy functional as
well as terms arising from XC-effects.
That is, evaluating $\Delta_{xc}(R)$ accurately is somewhat complicated. 
A strictly density-functional approach is discussed in Appendix B
of ref.~\cite{eos95}.

A simplified approximate procedure is as follows.
The core-core interaction can be written as a sum of
 monopole, dipole, quadrupole
and higher terms. In systems with free electrons (e.g., plasmas), only
the dipole term of the expansion is of importance due to screening effects. 
The frequency-dependent part of the dipole interactions
brings in the van der Waals (vW) type of contributions known as ``dispersion
 forces''.
These  have been discussed extensively within DFT~\cite{Maggs87,Ber2015vWals}.
While they are easily included in the NPA approach,
they are hard to include in the usual $N$-center DFT used in codes like the
VASP because of their
strong non-locality. One method used in $N$-center DFT calculations
is to decompose the $N$-center charge density into $N$ individual charge
distributions (each like an NPA). Using maximally
localized Wannier functions on each atomic center together with a
polarizability analysis for each center is one typical approach.
 The need for such procedures is an artifact of the
 usual $N$-center DFT approach
where even a fluid is represented as an average over an ensemble of
MD generated solids. The NPA scheme directly provides
 such single-center  distributions in the appropriate
form; hence dipole forces and vW effects are easily included when needed.

In practical WDM calculations where free electrons are present,
the  metallic interaction $V_{\rm ab}^{ff}(k)$, Eq.~\ref{ppot-f.eqn} is 
dominant. Then a fairly simple procedure may be sufficient for including
  core-core interactions when they become relevant.
For instance, given an atom `a' with an
argon-like core, and an atom `b' with a krypton-like core, we use 
the known Ar-Kr interaction given as a multipole expansion
 and screen it  using the free-electron
dielectric function of the WDM electrons at the appropriate 
free-electron density $\bar{ n}$ and $T$. 

We illustrated this using an example, namely,
a mixture of neutral  argon atoms Ar$^0$, together with
 argon ions Ar$^{Z_{i}+}$ with appropriate
integer ionizations $Z_i$.
In an argon plasma with, say, $\bar{Z}=0.3$
it is clear that most of the atoms ($\sim$ 70\%) are neutral
argon atoms, while some $\sim 30$\% are singly ionized Ar$^+$ ions.
Treating such a system needs the neutral Ar-Ar interactions
which are purely core-core interactions. We also have Ar-Ar$^+$  
as well as Ar$^+$-Ar$^+$ interactions, i.e., three
pair-potentials and three ion-ion PDFs associated with them. 
A single average-atom NPA or a naive $N$-atom DFT calculation
is quite inadequate for accurately estimating
physical properties of such a system.

In evaluating physical properties of the Ar, Ar$^+$ mixture, say,
 the self-diffusion coefficient,
there are two self-diffusion coefficients, and one inter-diffusion
coefficient as well. They are also constrained by the
compressibility sum rule. 
In such instances, the meaning of the self-diffusion coefficient
obtained from an $N$-center DFT calculation (e.g., a VASP calculation)
needs to be examined more carefully in the context of partitioning
the results to ascribe them to individual ionization species. 

Evaluations of core-core potentials for neutral argon atoms in a medium
without free electrons give results  close to parameterized
potentials similar to the Lennard-Jones(LJ) or more advanced 
rare-gas potentials. The presence of free electrons leads to screening 
of the core-core interaction. Thus, at the
LJ-level of approximation, $V^{cc}(k)$ for two neutrals immersed in
the appropriate electron gas is approximately given by
 $V^{LJ}(k)\{1+V_k\chi(k,T)\}$.

\begin{figure}[t]
\begin{center}
\includegraphics[width=\columnwidth]{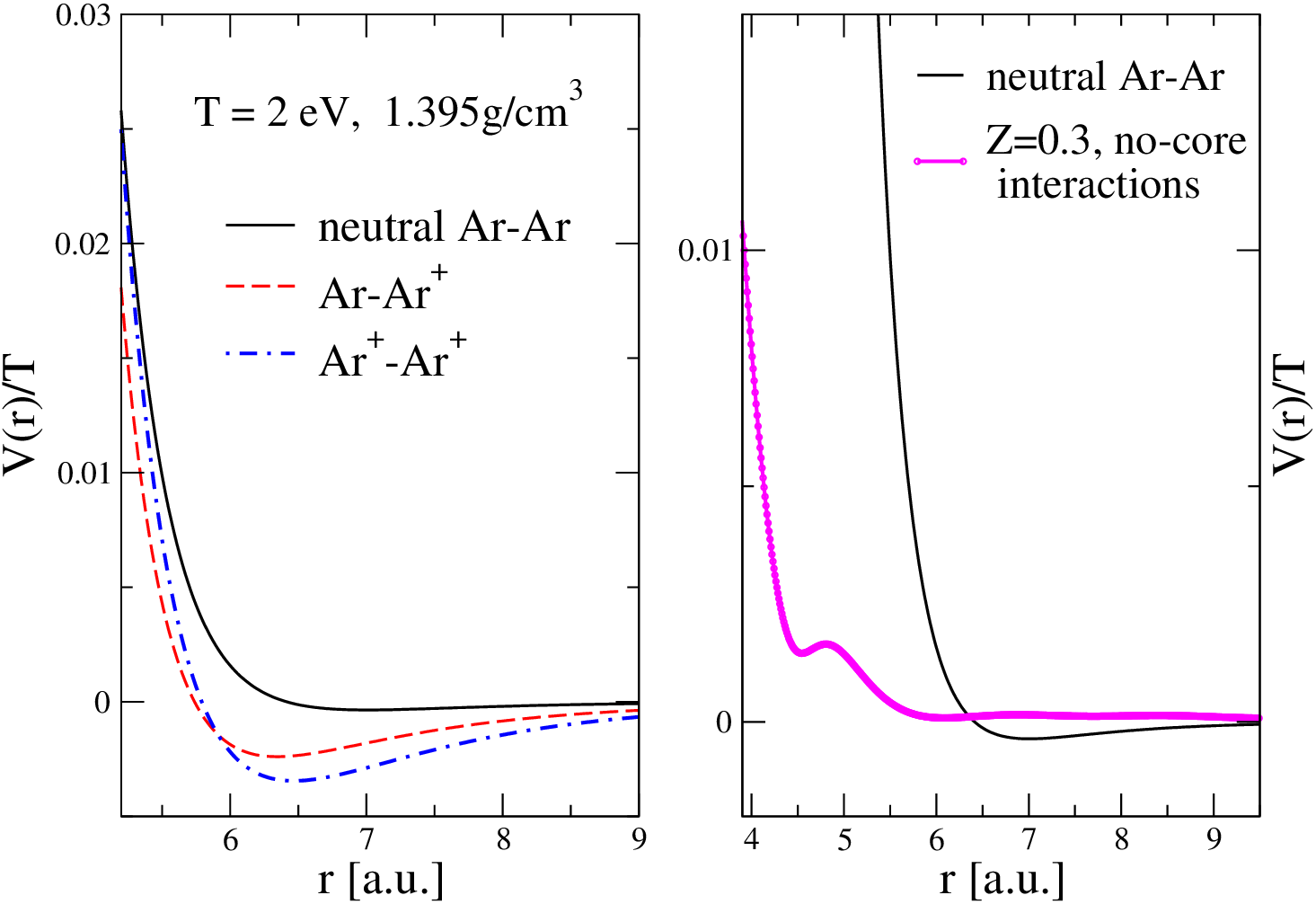}
\caption{
\label{argpots.fig}
(Online color)The screened pair-potentials Ar-Ar, Ar-Ar$^+$
Ar$^{+}$-Ar$^{+}$ of a two-component mixture of  argon at 2 eV. 
The single Average-Atom pair-potential $V_{\rm a,a}^{ff}(r)$, a=Ar$^{0.3+}$
is obtained (thick magenta curve) for
ions with $\bar{Z}$=0.3, without including core-core interactions}   
\end{center}
\end{figure}

For two charged ions, the major interaction
is via free electrons, and is given by the usual form, Eq.~\ref{ppot-f.eqn}.
The major part of the 
interaction between the ion and the neutral atom is the energy of
the induced dipole of the neutral atom interacting with the electric
field of the ion, moderated by screening. There is also
a shell-shell repulsion, giving rise to a modified LJ-like potential.
The polarizability and other parameters can be evaluated using the
bound electron densities of the Ar atom and the Ar$^+$ ion obtained
from the respective NPA calculations, or using known LJ
parametrizations. In Fig.~\ref{argpots.fig} we display the
three pair-potentials relevant to argon at normal density and
$T=2$ eV, when $\bar{Z}=0.3$, and when treated as a mixture. However,
if the system is treated as a single component fluid of
average atoms with a charge
of $\bar{Z}$=0.3, the resulting pair-potential evaluated
without including core-core corrections is found to be quite
different (right panel, Fig.~\ref{argpots.fig}) from those
inclusive of core corrections.

\subsection{The response function of the uniform electron fluid}
\label{uef.sec} 
The response function is itself a functional of the one-body electron
density and can be evaluated self-consistently within the calculation
by solving the NPA-Kohn Sham equation for just the central cavity but
{\it without} a nucleus. Then the free electron density pile up
$a_f(k)$ is entirely due to the cavity potential $V_{cav}(k)$ which
is the electrostatic potential of a weak spherical cavity.
Then Eq.~\ref{pseudo.eq} can be used in an inverse sense
to calculate $\chi(q,T)$ {\it in situ}. While such an approach is
useful as a control calculation, the following
direct procedure has been implemented in our codes.
   
The interacting electron gas response function used in these calculations
includes a local-field factor (LFC) chosen to satisfy the finite temperature
electron-gas compressibility sum rule.
\begin{eqnarray}
\label{resp.eq}
\chi(k,T)&=&\chi_{ee}(k,T_e)= \frac{\chi_0(k,T_e)}{1-V_k(1-G_k)\chi_0(k,T_e)},\\
\label{lfc.eq}
G_k &=& (1-\kappa_0/\kappa)(k/k_\text{TF})^2; \quad V_k =4\pi/k^2,\\
\label{ktf.eq}
k_{\text{TF}}/k_{F}&=&\{(4/\pi) \alpha r_s)\}^{1/2};\quad \alpha=(4/9\pi)^{1/3},\\
\label{vii.eq}
 V_{ii}(k) &=& Z^2V_k + |U_{ei}(k)|^2\chi_{ee}(k,T_e).
\end{eqnarray}
Here $\chi_0$ is the finite-$T$ Lindhard function, $V_k$ is the bare Coulomb
potential, and $G_k$ is a local-field correction (LFC). The finite-$T$
compressibility sum rule for electrons is satisfied since $\kappa_0$ and
$\kappa$ are the non-interacting and interacting electron compressibilities
respectively, at the temperature $T$, with  $\kappa$ matched to the
 $F_{xc}(T)$ used in the KS
calculation. In Eq.~\ref{ktf.eq}, $k_\text{TF}$ appearing in the LFC is the
Thomas-Fermi wavevector. We use a $G_k$ evaluated at $k\to 0$ for all $k$
instead of the more general $k$-dependent form (e.g., Eq.~50  in
Ref.~\cite{PDWXC}) since the $k$-dispersion in $G_k$ has negligible
effect for the WDMs of this study.

\section{The NPA pair-potential approach compared to methods based on
 multi-center potentials}
\label{pair-multi.sec}
The pair-potential $V_{\rm ab}(R)$ is normally understood as
 the energy of the system with two `atoms', or `ions' as the case may be (but
 referred to here as `atoms') and denoted by `a' and `b' and
 held at a separation  $R$ , compared to the limit $R\to\infty$
  when there is no interaction  between them. 
This has a clear meaning if the  system is in a vacuum and
 at sufficiently low $T$ such that `a' and `b' remain as compact objects.
 Thus the  `bond energy' of two hydrogen atoms calculated using standard
 quantum chemistry methods, at $T=0$, has no ambiguity. However, in an
 atomic or molecular fluid, or in a WDM, the medium contains other atoms
and even free electrons, with $T\ne 0$.

So one needs to include the effect of neighbouring
atoms that are in the medium.  This is exactly the problem systematically
 addressed by many-body theories  like DFT. The effect of the medium is a
 functional of the  {\it one-body densities} of the components that make up
 the medium where the two atoms `a' and `b' are placed in. Instead of
 constructing the necessary functionals in a systematic way, keeping in mind
 that they should be one-body functionals,
semi-empirical potentials (fitted to data bases etc.) have deployed
 multi-center models mimicking bonds, bond angles, their torsional properties
 and so forth, based on  preconceived `chemical bonding' pictures.
 If this `fitting'
 had been directed  to constructing just the
 electron XC-functional, then we have the effort found in quantum chemistry,
  colourfully described as constructing a ``Jacob's Ladder'',
where increasingly sophisticated  electron XC-functionals are constructed
 for electronic  systems placed in the  potential
of a finite number of nuclei, usually held fixed.

In the following we illustrate the DFT-NPA approach of using pair-potentials
and XC-functionals by comparing it with the  Stillinger-Weber (SW)
 model as a well-known example of a multi-center potential
useful for tetrahedral materials, be they in solid, liquid or even WDM states
like molten silicon or liquid carbon.

\subsection{The $N$-body energy in the Stillinger Weber model}
We briefly recount the details of the SW model for the convenience of the
reader. The quantity they model is the potential-energy function $\phi$
for $N$ identical interacting particles. SW write the potential for Si as:
\begin{equation}
\label{SW.eqn}
\phi(1,2,\ldots,N)_{\rm sw}=\sum_{i<j}V_2(i,j)+\sum_{i<j<k}V_3(i,j,k). 
\end{equation}
A one-body term is not displayed as it represents external potentials.
SW truncate their expansion in  third order. The two-body term
is written as the sum of a repulsive term $V^{re}$, and an
 attractive term $V^{at}$.
An energy scale
$\epsilon$ and a length scale $\sigma$ are introduced as in the
 Lennard-Jones potential.
The three body term should possess translational and rotational symmetry,
 and contains 
the bond angle $\theta_{jik}$.
The three-body terms $V_3$ sums over the neighbours $k$ of the pair
$i,j$. 
\begin{eqnarray}
\label{sw23.eqn} 
V_2{i,j}&=&V_2(r_{ij})=V^{re}(r_{ij})-V^{at}(r_{ij})\\
V_3(i,j,k)&=&U(r_{ij})U(r_{ik})(\cos\theta_{jik}+1/3)^2\\
          &+&U(r_{ji})U(r_{jk})(\cos\theta_{ijk}+1/3)^2\\
          &+&U(r_{ki})U(r_{kj})(\cos\theta_{ikj}+1/3)^2\\
V^{re}(r)&=&A\epsilon(B/x^p)\exp\{1/(x-a)\}\\
V^{at}(r)&=&A\epsilon(1/x^q)\exp\{1/(x-a)\} \\
     U(r)&=&\sqrt(\lambda\epsilon)\exp\{\gamma/(x-a)\},\;x=r/\sigma .      
\end{eqnarray}
Here the angular part, with the $\cos\theta_{jik}$ function contributes 
strongly in favour of pairs of bonds arising from the atom `$i$' that conform to
the tetrahedral configuration common in C, Si etc., at normal densities
 and temperatures.
That is, structural features are built into the potential assuming the validity
of specific structures. The sum of contributions to the three-body forces
from an ensemble of SW-atoms tends to zero as the structure tends to tetrahedral
bonding, while other possible structures become destabilized by an unfavourable
three-body contribution. It is standard-writ in classical MD simulations that
 `three-body terms are necessary' to stabilize the diamond structure, but true
only in the sense that LJ-type short-ranged pair potentials fail. The SW model
 contains seven fit parameters, of which $q$ is set to zero
 in Eq.~\ref{sw23.eqn}. All three pair-functions that make up the SW potential
vanish at $r=\sigma a$, with $\sigma= 3.959$ a.u., $a=1.80$ and
$\epsilon=2.1682$ eV, i.e., 25,160K for Si~\cite{SW85}. Currently, there is a
plethora of SW potentials with somewhat different parameter fits but the basic
idea has been to ``fit a  potential'' to structure data incorporating 
models of bond stretching, bond bending etc., rather than using one-body
DFT functionals.
 
\begin{figure}[t]
\begin{center}
\includegraphics[width=\columnwidth]{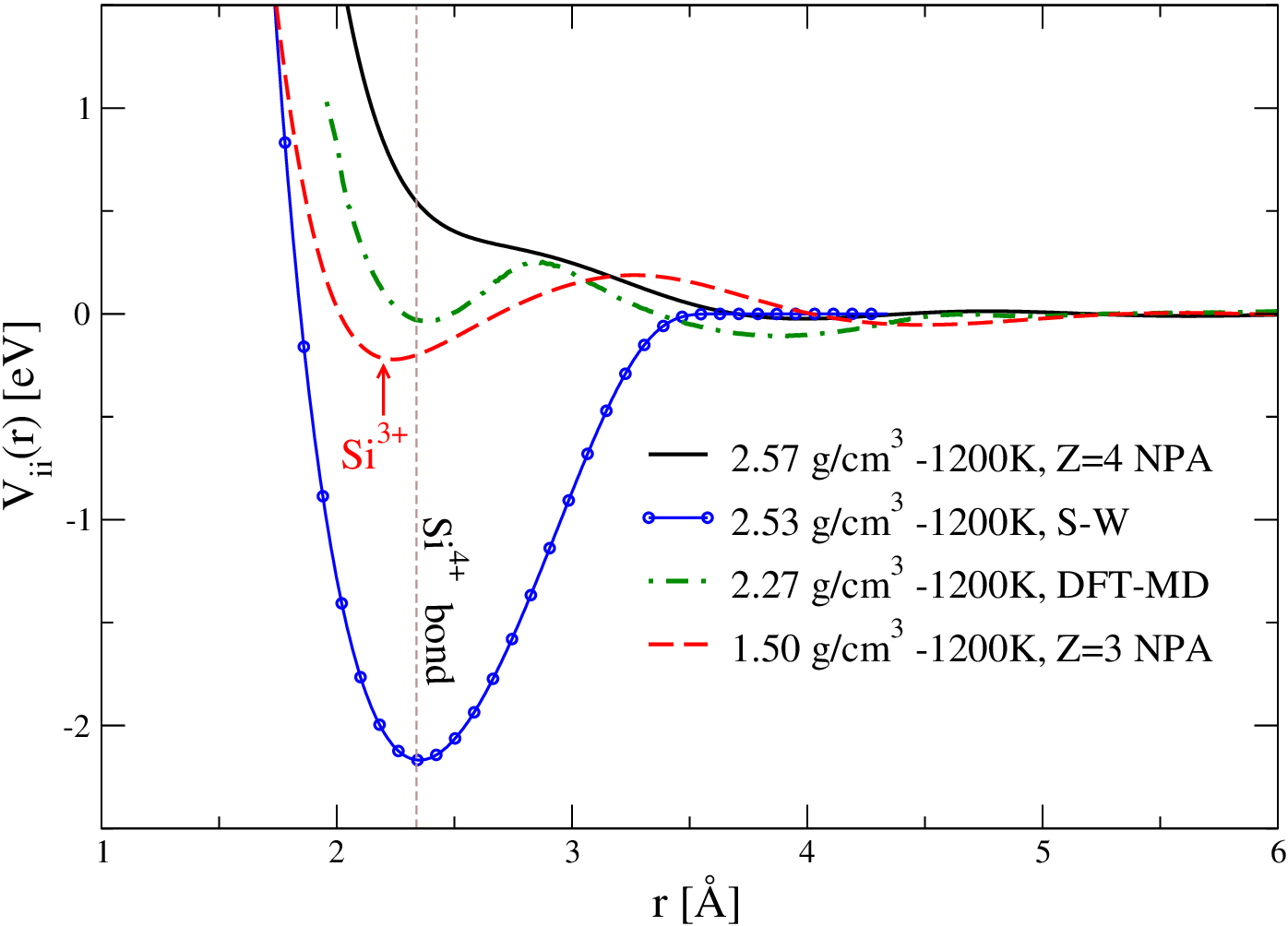}
\caption{
\label{v12SW-npa.fig}(color online) The pair-potentials from the NPA, from
a VASP calculation with 216 atoms using the SCAN XC-functional for
supercooled liquid silicon at 1200K, and at the metastable low density of 2.27g/cm$^3$
are shown. The normal
melt density is higher, being at 2.57 g/cm$^3$ prevails above 1683K. The pair-potential
for the unstable higher density liquid from NPA at 1200K is also shown.
The Stillinger-Weber pair-potential for liquid Si at 1200K is
 contrasted with the other potentials. We also display the NPA potential for
 expanded molten Si at 1.5g/cm$^3$ where a low-ionization state Si$^{3+}$
 occurs. The Si-Si nearest-neighbour ``bond'' distance is shortened
 for Si$^3+$ ions, compared to the Si$^{4+}$ pairs.}
\end{center}
\end{figure}

In fig.~\ref{v12SW-npa.fig} we compare the SW pair potential to the
NPA pair-potential and the DFT-MD potential extracted from  a simulation of
216 particles~\cite{Remsing17}, using the SCAN functional~\cite{SCAN13} that
 is most appropriate
 for systems with covalent bonding. The very demanding case of supercooled
molten silicon at 1200K  has been used for a comparison of the NPA and DFT-MD
VASP simulations. We choose this case as a detailed account
of NPA potentials for molten supercooled Si is given
in Ref.~\cite{cdwSi20} and in the supplemental material associated with it.

A `built-in' feature of the pair part of the SW-potential is its deep negative
energy near the Si$^{4+}$ bond length in the solid. 
Such a deep feature is not found in  DFT-VASP
or in the DFT-NPA potentials. Unlike in SW containing only classical ions,
the NPA and VASP calculations treat silicon as a
 two-component system with an ion subsystem and an electron subsystem. The
 stability of the system arises from the ionic interactions as well as
 a volume energy 
associated with the electron fluid. Hence the DFT models have a
 comparatively shallow
ion-ion potential, while the SW potential has to be deeper to include the
energy of the electron subsystem as a ``bond energy'' component. Furthermore,
while a very short-ranged pair-potential alone is insufficient to `stabilize'
the one-component tetrahedral structure, a long-ranged pair-potential
 inclusive of the
`electron gas' contribution to the total energy covers all the
  possible electron-ion
structures that are quantum mechanically possible. As already noted in
discussing the potentials and PDFs of Al, Ar, hydrogen, and in previous
 publications on C, Si~\cite{DWP-carb90}, Al~\cite{cdw-aers83}, the energy
minima in the Friedel oscillations of the ion-ion pair potential, together
with the volume dependent energy effects of the electron fluid,
 were seen to correlate closely  with the positions of the higher-order
 neighbours beyond the nearest neighbour.

\begin{figure}[t]
\begin{center}
\includegraphics[width=\columnwidth]{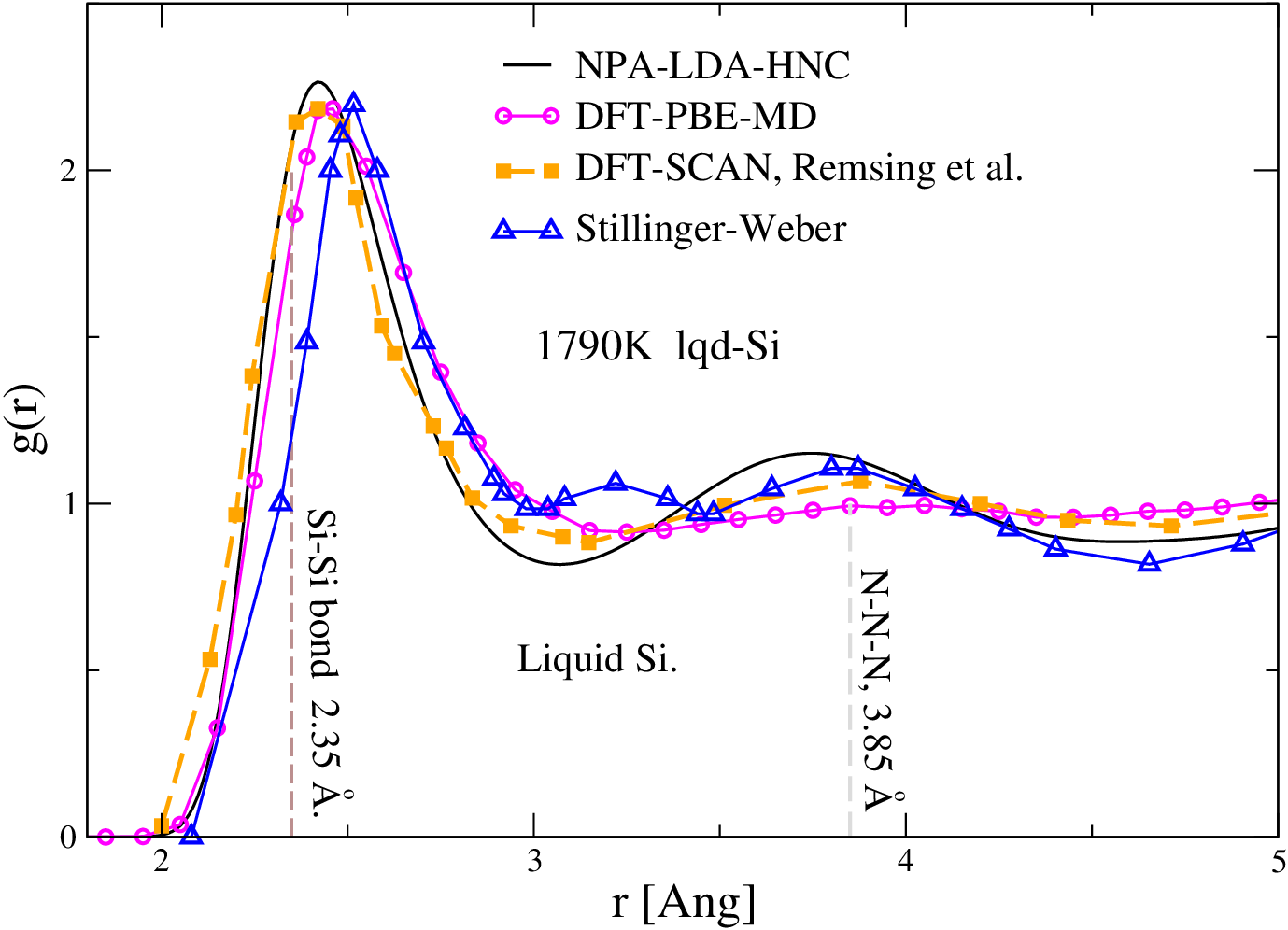}
\caption{
\label{grSi.fig}
(Online color) The PDFs for molten Si slightly above the melting point
calculated by several methods is displayed. Two DFT-MD calculations
 using the VASP,
code, and with the SCAN and PBE XC-functionals show  sensitivity
 to the  functionals used in the
 216 atom calculation. The NPA, using only
one Si atom has a much smoother electron distribution and obtains
 comparable results with an LDA XC-functional. The indicated 
nearest-neighbour (N-N)  and N-N-N distances are for solid Si.
The S-W model shows a spurious hump near 3-3.5 \AA\ . 
}
\end{center}
\end{figure}

A comparison of the Si-Si $g(r)$ obtained from various calculations is given in
Fig.~\ref{grSi.fig}.  The PDFs from NPA calculations, DFT-MD calculations
using the SCAN XC-functional, as well as the PBE~\cite{PBE96} XC-functional are
shown, together with a typical classical MD simulation for the SW potential
~\cite{YuSW96}. The SW potential is chosen here for comparison as it is
superior to the Tersoff potential and other EMA potentials for modeling
liquid-Si~\cite{Cook-Si93}.

The differences between the one-center NPA with LDA, 216-atom
 DFT-MD with PBE, and with SCAN are similar in magnitude; comparison with experiment
shows that none is superior to the other~\cite{cdwSi20}. The one-center
 electron density  distribution used in the NPA
is very simple and smooth compared to the 216-center $n(r)$ used in the VASP
calculation. In fact, the LDA-XC is generally found to work efficiently and
accurately for the NPA. Furthermore, the NPA calculation with the
 hyper-netted-chain (HNC) equation is computationally much faster and
 cheaper than the SW simulation, while also being a first-principles
 DFT calculation.

\subsection{The ion-ion $N$-body corrections and the NPA model}

It is of interest to see how the pair-potential used in the NPA brings in the
three-body energy and such `multi-center'  terms via the ion-ion XC functional
used in the NPA approach. Of course, if classical MD is used with the NPA
pair-potentials~\cite{Nadin88,HarbourDSF18,Stanek21}, then the ion-ion
correlations are automatically built up by the long-range pair-potentials.
Such simulations yield the $g_{ii}(r)$, but the calculation of the total
free energy requires contributions from the electron subsystem to the
total energy, as discussed in Ref.~\cite{eos95}. 
In using the SW potential, the $g_{ii}(r)$ and the SW-potential
are sufficient to determine the total energy as there is no electron subsystem.

Here we consider the question of many-ion effects using the theory of integral 
equations for ion correlations. In contrast to the SW-potential, the NPA pair
potentials do not include any pre-assigned structural characteristics except for
the effect of the average ion density imposed by the Wigner-Seitz sphere.
 SW-type pre-assigned structures are usually valid only in a specific
 regime of
$\bar{Z}$ (e.g., carbon with a valance of four), $\bar{\rho}$.
 The pair-potentials
and  the associated XC-functionals of the NPA approach {\it generate}
the appropriate lowest energy structure where
the local ordering may be tetrahedral, face-centered etc., or a thermodynamic
mixture of many structures with no predominant structure. 

We consider a uniform fluid consisting of atoms of one type for simplicity.
 The DFT
Eq.~\ref{KS-ion.eqn} for the ion subsytem solves for the minimum energy
 ion distribution $\rho(r)$ and gives the form:
\begin{equation}
\rho(r)=\bar{\rho}g(r)=\bar{\rho}\exp\{-V_I(r)/T\}.
\end{equation}
where $V_I(r)$ is the density-functional potential felt by an ion $I$ at the
 radial location $r$, in the presence of an ion at the origin. 
 This  potential  can be  written within DFT as:
\begin{eqnarray}
\label{ionpot.eqn}
V_I(r)&=&V_{0I}(r)+V(\rm{mean\ field})+V_{ii}^{xc}(r)\\
V_{ii}^{xc}&=&V_{ii}^c(r)=\delta F_{ii}^{xc}[\rho(r),n(r)]/\delta\rho(r).
\end{eqnarray}
The first term on the r.h.s., $V_{0I}$ is the ion-ion pair potential between
 the central ion at the origin
and the ion $I$ at $r$, viz., the NPA pair-potential.
 Any arbitrary ion is denoted by the lower-case $i$. 
The ion $I$ is also subject to the self-consistent average potential
at $r$ from all the field ions in the medium, i.e.,
 from the  density
 $\rho(r)=\bar{\rho}g(r)$. The effect of the
 electron distribution $n(r)$ has to be included in the mean-field potential. 
The relevant electron density is that given by the  solution
 of the Kohn-Sham equation, viz., Eq.~\ref{KS-basic.eqn}, coupled to the ion density.
Then the mean-field potential is just
 the Poisson potential.
 when evaluated at the
level of a monopole expansion, i.e., neglecting core effects and pseudopotential
form factors.
Defining the convolution integral by
\begin{equation}
f*g=\int d\vec{r}_1f(r_1)g(|\vec{r}-\vec{r}_1|),
\end{equation}
the Poisson potential is given by 
\begin{equation}
\bar{Z}\{\rho-n\}*\{V\},\; V(r)=\bar{Z}/r.
\end{equation}
Terms not contained in the mean-field are in the ionic-XC term $V_{ii}^{xc}(r)$. 
We also neglect electron-ion XC effects and use the Born-Oppenheimer
approximation (see Ref.~\cite{ilciacco93}).
Since ions are classical, there is no exchange, and this is purely
a correlation term that can be calculated using classical statistical
mechanics. 

The ion correlation free energy is formally given as a coupling-constant
integration on the $g_{ii}(r)$~\cite{eos95}, just as for the electron XC-functional.
But $g_{ii}(r)$ is initially unknown. An LDA  $V_{ii}^c(r)$ based on the
 classical correlation energy of the uniform classical Coulomb fluid, similar to
the LDA for electron correlations fails, as reported
in Ref.~\cite{DWP1}. It is found that $V_{ii}^c(r)$ is
highly nonlocal and even gradient expansions fail. Hence we exploit the
relationship of the XC-functional with the corresponding pair-distribution
functions as follows.    

The  three-body and higher correlations
neglected in the mean-field potential are included via
the Ornstein-Zernike relation which expresses the total correlation function
$h(r)=g(r)-1$ in terms of the direct correlation function $c(r)$. 
We write the Ornstein-Zernike equation as:
\begin{equation}
h_{ii}(r_{12})=c_{ii}(r_{12})+ \bar{\rho} c_{ii}(r_{13})*h_{ii}(r_{31}). 
\end{equation} 

This equation is a simple algebraic equation in $k$-space for fluids of
uniform-density. It brings in the interaction of the two atoms $1,2$
with all possible `third atoms' at the location $\vec{r}_3$ which
runs over all space.    
The explicit correlation corrections are selected to be the sum of hyper-netted-chain
(HNC) diagrams excluding mean-field terms that have already been
taken into account in Eq.~\ref{ionpot.eqn}. That is, noting that $r$=$|\vec{r}_{12}|$
in the uniform case, and using the short-ranged direct correlation function
$\tilde{c}_{ii}(r_{12})$ we have
\begin{eqnarray}
\label{ion-cor.eqn}
V_{ii}^c(r)/T &=&-\bar{\rho }\,\tilde{c}_{ii}*h_{ii} \\
\tilde{c}_{ii}(r_) &=& c_{ii}(r)+V_I(r)/T.  
\end{eqnarray}
Classical correlations not captured by the HNC sum are not easily evaluated
and are bundled into a `bridge diagram'  evaluated using
hard-sphere models~\cite{LFA83}.
 
In actual NPA calculations where a large correlation sphere of radius $R_c$ is used,
all quantities are referred to the limit $r\to R_c$. This is true for all
XC-potentials as well as other quantities like chemical potentials.
That is, $V_{ss'}^{xc}(r)$ are effective values $V_{ss'}^{xc}-V_{ss'}^{xc}(R_c)$. 
\subsection{Expressions for total Free energy in the NPA model}
 
For comparison with expressions to be given below for effective medium theories,
we discuss the total Helmholtz free energy $F$ of a fluid of neutral pesudoatoms.
The evaluation of the free energy is more complicated than that of the
 internal energy $E_{tot}$, as calculations of the entropy contribution to $F$ in DFT
 is not conceptually straightforward especially in dealing with atomic bound states
with partial occupancies. The  total free energy per atom for a given
ionic configuration (expressed as the density distribution $\rho(r)=\bar{\rho}g(r)$
around a given  nucleus) is schematically summarized below, while the full
 expressions may be
found in Refs.~\cite{Pe-Be} and \cite{eos95}.
\begin{eqnarray}
\label{FEn.eqn}
F&=&F_{id}+F_0+F_{em}+ F_{12}-E^0_{at}\\
F_0&=&\bar{Z}f^h(\bar{n},T)=\bar{Z}(f_{k.e}+f_{xc})\\
F_{em}&=&\Delta F_0+\Delta V_{ei}(\Delta n)+\Delta V_{ee}(\Delta n \Delta n')\\
F_{12}   &=&(1/2)\sum'_j V_{12}(R_j)+\mbox{cavity \ corrections}.
\end{eqnarray}
The energy of an isolated atom at $T=0$ in its ground state, $E^0_{at}$,
is used as the  reference  energy.
Here $F_{id}$ is the classical ideal-gas energy of the non-interacting ion subsystem
per atom. Each atom contributes $\bar{Z}$ free electrons to the electron
fluid.   $F_0$ is the free energy of $\bar{Z}$ electrons in a uniform electron gas,
$f^h$ the homogeneous free energy/electron, containing
 a kinetic energy component and an XC-component. The
``embedding free energy'' of a neutral pseudo atom is $F_{em}$. 
This is  defined as the difference between the free energy of the electron
fluid of mean density $\bar{n}$ with and without the NPA. It consists of a
 correction to $F_0$, the Coulomb interaction of the pseudo-ion with the displaced
 electrons, as well as the correction to the electron-electron repulsion energy
 due to the displaced electrons.

The $F_{12}$ term contains the free energy contributed
through the ion-ion pair-distribution function and the pair-potential. The many-atom
correlations dependent on the `environment' are included here via the ion-ion PDF. 
Hence this is effectively the pair-energy and the correlation corrections to the
free energy of  the classical ions. That is,
in more conventional language, the  `bonding energy' of two  ions immersed
in the fluid with average densities $\bar{n}$ and $\bar{\rho}$, inclusive of
 the effect
of their higher-order neighbours brought in via the secondary peaks of the PDFs. 

While $F_0$ is a smooth function of the
electron density, sharp jumps may occur in $F_{12}$, and to a  lesser extent in
$F_{em}$, signaling the onset of ionization changes, structural changes or
 phase transitions.
Example of such discontinuities at liquid-liquid phase transitions (LPT)
may be found in Ref.~\cite{cdwSi20} for liquid Si, while similar transitions
have been noted in theoretical studies of liquid
 carbon as well~\cite{glosli99,CPP-carb18}. Interestingly, a molten-carbon  LPT
 was predicted in Ref.~\cite{glosli99} using an empirical bond-order
multi-center carbon potential, but this was not confirmed when DFT-MD simulations
were done using the PBE functional~\cite{WuLPT02}

\section{effective medium potentials and NPA potentials}
\label{efm.sec}
Unlike the purely classical SW model and effective medium approach (EMA), 
embedded-atom models (EAM) pay attention to
 the existence of an electron subsystem, and begin by using 
approximations to $T=0$  DFT to include the
``effect of the local environment'' of an ion in the medium. 
The Finnis-Sinclair potentials~\cite{FinisSinc84}
 used in metallurgical applications
is based on a second-moment approximation to tight-binding but reduces
to an effective-medium type approach.
A  DFT type discussion is used in the initial theory of effective media to
justify the use of a form of the total energy which is then
parameterized using
a variety of empirical models containing fit parameters.
Hence systematic  generalizations for $T\sim E_F$ or higher $T$
are not available although low-$T$ fitted forms exist. Here
we examine the more systematic foundations which are, however,
of little use in evaluating many of the current
models which should be regarded as sophisticated empirical fits to data bases.

The total
energy of the electron-ion system is written in EMA as:
\begin{equation}
\label{ema1.eqn}
E_{tot}=\sum_i F\{n_i(\vec{R}_i)\}+(1/2)\sum_{i,j}\phi(\vec{R}_i-\vec{R}_j).
\end{equation}
Here $F\{n_i(\vec{R}_i)\}$ is a function of the electron density, 
with $n_i(\vec{R}_i)$ being electron densities at atomic sites $\vec{R}_i$.
Also, $\phi(\vec{R}_i-\vec{R}_j)$ is a pair-potential mainly used to
model the repulsive interactions among the atoms. The atomic densities are
associated with an embedding energy which is taken to be a function of the
local average one-body electron density $\bar{n}$, following DFT.
The embedding energy at $T=0$ as a function of the electron density
has been tabulated by various authors, e.g., Stott {\it et al}~\cite{StottZaremba85}.
 A superposition approximation
is introduced, and the total energy is finally expressed as a sum of terms
for the isolated-atom energies plus the embedding energy in the homogeneous
electron gas, and a gradient expansion in the electron density is used
to allow for the density changes at atomic sites. No attempt has
been made to take advantage of ion-DFT using an ion-correlation functional.

However, this approach turns out to be inadequate in many ways, yielding
wrong elastic constants etc~\cite{DawBaskes84}. The
embedded atom model (EAM) was a numerical improvement on the EMA. The EAM attempts
to merely use the structure of the energy expression for fitting
to parameterized forms similar to it. Below  we consider a more
sophisticated  form known as the `glue model'~\cite{Ercolessi94}. 

The parameter-free first-principles
 NPA pair-potential for Aluminum
is compared with one of the original force-matched Al `glue' potentials that uses
some 40 fit parameters and some 32 constraints.
The `glue' model uses the form:
\begin{equation}
\label{glue.eqn}
E_{tot}=\sum_i U\{\sum_jn(r_{ij})\}+(1/2)\sum_{i,j}\phi(r_{ij}).
\end{equation}
Here, instead of the  electron gas term used in the EMA, or in the NPA as
given by Eq.~\ref{FEn.eqn},  a `glue' function $U(n)$
is used, while $\phi(r)$ denotes the pair-potential. Hence, this aluminum potential
 depends on specifying three functions, $\phi(r), n(r)$ and $U(n(r))$ that contain
fit parameters and constraints. Here again, simplification of the
problem that could be achieved using a DFT ion-correlation functional
to include many-ion effects is not invoked. 
 The forces obtained from this potential (defined in terms of parameters contained
in polynomials or Pad\'e forms or other fit functions)
 are matched (by adjusting the fit parameters)
to those from  first principles calculations for a large variety of atomic
 configurations and physical situations (solids, liquids, clusters, surfaces,
 defects etc.), including those at finite $T$. The Al potential of Ercolessi {\it et al}.
 has been fitted to liquid Al  simulations at 1000K and 2200K (i.e., $T/E_F=0.016$). 
%0.1896 eV 

The NPA approach has no fit parameters,  and finds
 WDM states of Aluminum to  be  `easy' systems
for successful theoretical predictions. However,
Al is regarded as as a `difficult case' for effective medium  methods. 
Many of the EAM potentials give poor predictions of  aluminum melting temperatures,
vacancy migration  energies, diffusion constants, stacking fault energies,
surface energies, phonon spectra and so forth. The NPA potentials have been tested
mainly in WDM situations where they are in excellent agreement with DFT-MD
calculations of EOS properties, PDFs, as well as for transport properties like
the electrical conductivity~\cite{cond3-17} and diffusion constants~\cite{Stanek21}.
 Unlike the EAM, the NPA provides pseudopotentials
and eigenfunctions  needed for many other calculations, e.g., linear transport
coefficients, line broadening, XRTS spectroscopy
and energy relaxation~\cite{elrDW01}.

A key difference between these fitted-potential approaches, and that of the NPA, is
the inclusion of environment-specific features in the NPA, be it for a liquid
 or a solid, via the pair-distribution function $g(r)$  of the ionic structure
 appearing in the ion-correlation functional of the NPA, viz., Eq.~\ref{ion-cor.eqn}.
The needed $g(r)$ is  generated {\it in situ} for fluids. In the case
of specific crystal structures at low $T$, one can use a known explicit form,
as in Perrot's Be calculations~\cite{Pe-Be}, or the phonon calculations of
Harbour {\it et al}~\cite{HarbEOSPhn17} using the NPA.

\begin{figure}[t]
\begin{center}
\includegraphics[width=\columnwidth]{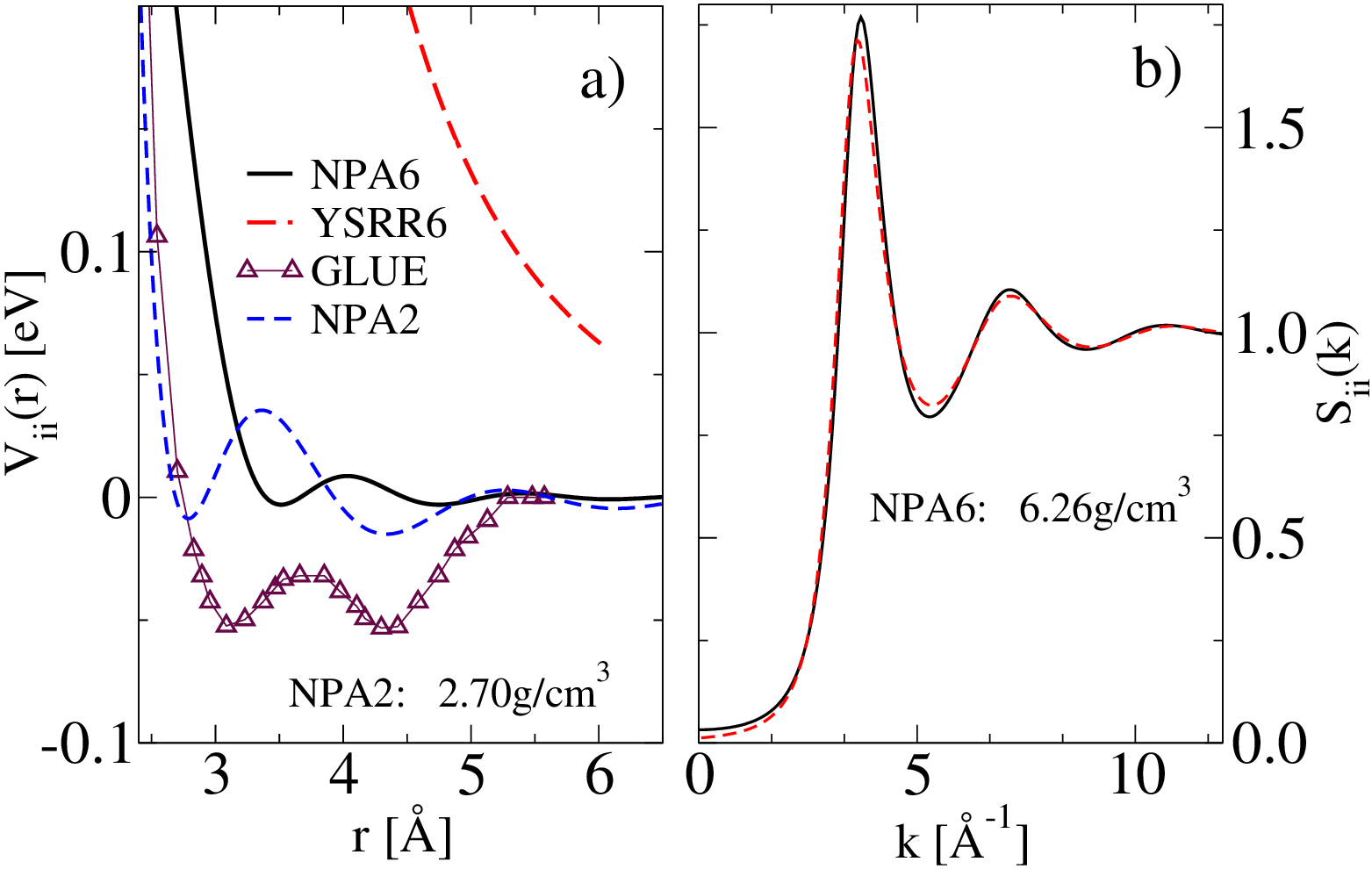}
\caption{
\label{glue.fig}
(Online color) Panel (a) displays the first-principles pair potential
form NPA labeled NPA2 (blue dashed line) at $T$=2200K and the normal Al density
of 2.7g/cm$^3$, as well as the pair part of the `glue' potential (triangles) of
 Ercolessi  {\it et al}. NPA6 is the pair-potential for compressed Al at a density
of 6.26 g/cm$^3$ and $T$=1.75 eV studied experimentally by
 Fletcher {\it et al}~\cite{Flet-Al-15}. A
model potential (a Yukawa potential added to an {\it ad hoc} short-ranged repulsive
 potential) labeled YSRR6 used by Fletcher {\it et al}. 
is also shown (red dashed lines).  In panel (b) the
structure factors from the NPA6 potential and the YSRR6 potential are displayed.
The YSRR6 potential reproduces the experimental $S(k)$ as it has been fitted to its
$g(r)$. The NPA6 is a first principles calculation with no fit
parameters~\cite{xrt-Harb16}. 
}
\end{center}
\end{figure}

In Fig.~\ref{glue.fig} we compare the NPA Al-Al pair potential with that of the
`glue potential' of Ref.~\cite{Ercolessi94}. In panel (a) we display the NPA pair
potential, labeled NPA2, for liquid aluminum at the normal density of 2.7 g/cm$^3$
 and $T$=2200K. The
pair part of the glue potential is also shown (curve with triangles). Unlike in the
case of the SW potential, the glue potential has Friedel like oscillations, although 
not as long-ranged as in the NPA2 potential. The force-matching methods do not
usually  have the accuracy to recover the weaker Friedel oscillations which
decay as $1/r^3$.  The additional
term of the glue potential, i.e, $U(n)$ (not displayed in the figure)
 brings in contributions to the total energy
that are also included in the NPA via the electron-fluid term and the XC-correlation
energies of ions as well as electrons, as indicated in Eq.~\ref{FEn.eqn}. 
The ion-correlation energy is structure
dependent, as indicated in Eq.~\ref{ion-cor.eqn}.
 
In Fig.~\ref{glue.fig} we also consider the case of Aluminum at
a density of 6.264 g/cm$^3$, and at a temperature of 1.75 eV. This corresponds to 
$T/E_F=0.086$ as $\bar{Z}=3$ for Aluminum even at this temperature and compression.
The compression drives up the Fermi energy and the effective temperature $T/E_F$
remains low and the electrons are highly degenerate.
The system has been studied experimentally and theoretically by Fletcher
{\it et al}.~\cite{Flet-Al-15} using a physically motivated {\it ad hoc} 
potential (YSRR6)~\cite{Wunsch09} shown as a red dashed line. The YSRR potential is
made up of a Yukawa (Y) potential and a short-ranged repulsive (SRR) potential. 
The YSRR6 potential at the density of 6.24g/cm$^3$ had been fitted
 to the DFT-MD $g(r)$, and the corresponding $S(k)$ is shown as a red-dashed
 line in panel (b). 

The first-principles potential from
 the NPA at this density and temperature is labeled NPA6.
 It is very different (black solid line) from the
YSRR potential, but gives excellent agreement  with the MD-DFT $S(k)$ as well
 as with the XRTS data  of Fletcher, without any fitting.
 That many potentials produce only a few crystal structures or very
similar liquid structures, and hence the inverse problem is open to much
 uncertainty is well established~\cite{chenlai92}. However, the
 spurious candidates for the potential can be  eliminated
by their failure to predict other physical properties.
The YSRR potential predicts a very low compressibility,
very stiff phonon spectra, as well as unrealistic electrical conductivities, as
discussed in detail by Harbour {\it et al}~\cite{xrt-Harb16}. These emphasize the fact
that fitting to a structural feature (or even several) is no guarantee of having
approximated the physically correct potential. 
 
\section{Discussion}
The idea of using an ion correlation functional for treating ion-ion many body
effects in ion-electron systems was proposed as early as 1982~\cite{DWP1}.
 It was motivated by the success of using an electron
XC-functional to treat electron many-body effects. The method  leads to
a theory based on a pseudoatom, but it has not found significant
adoption, partly because materials science has been mostly concerned with treating
 materials-science
problems involving defects, dislocations etc., or
 with complex materials. Similarly, the idea of an ion correlation functional is
of not much use for molecules and other finite systems used in quantum chemistry.

However, there is little excuse for not using pseudo-atom approaches for uniform
 systems at finite-$T$ as the conceptual and computational advantages are very
 significant.

 A common  misconception, partly arsing from the `chemical picture' of bonding
 as the basis of complex correlations,
 as well as the wish to subsume the electron subsystem in bonds, has lead to
 the belief that the NPA ``one-atom'' approaches do not contain,
and cannot include, many-ion effects associated with the local structure
 of the medium. Here we have given examples showing how the NPA calculations
 incorporate such effects,  and clarified the kinship of the NPA approach and
 its energy functionals to those used in  effective medium models. The NPA remains
 a strictly first-principles DFT approach, while the EMA
 approaches have now become largely spring boards  for data fitting.
 The ability of the NPA to study subtle
effects like liquid-liquid phase transitions in complex materials like liquid Si
 and liquid C has been demonstrated and these have been reviewed in the
 context of how ion correlations are included in the NPA pair-potential scheme.
 However, much of the success of the NPA depends on the existence of free
electrons in the system that weaken the interparticle interactions. Additional
steps are needed in treating weakly ionized systems and transition metals
at very low temperatures.

Furthermore, although much effort has been directed to constructing
 kinetic energy functionals in the hope of simplifying DFT calculations, such
 calculations for $N$-atom systems
will also not directly yield useful `one-atom' properties like $\bar{Z}$, fractional
 compositions of ionic species,
pseudopotentials and pair-potentials that are valuable intermediates
 for additional calculations of
physical properties.
In contrast, the  NPA method provides such quantities directly,
 and may also easily include effects like van der Waals energies, and quantum
 nuclear corrections  in its potentials.    

\section{conclusion}
The NPA method has been successfully applied to a variety of
 warm-dense-matter systems, as well as
to a number of solid-state systems. The detailed study of Si for supercooled
 and hot liquid silicon has shown that its accuracy is similar to DFT-MD
 simulations that may themselves  differ from each other at the level of
 the electron XC-functionals used. The NPA, with its
simpler one-body electron densities allows the use of the local density approximation
in the XC functional. The NPA yields one-atom properties like the ionization state,
fractional compositions of ionic species,  pseudopotentials
etc.,  not directly available from the $N$-atom VASP-type simulations.
 On the other hand, the NPA being
a static DFT theory, does  not provide the complex bonding structure that
 may prevail transiently in short-time scales. The thermodynamic
properties of the system do not depend on such transient effects, and the
NPA deals only with long-time averages. 
\newline

{\bf DATA AVAILABILITY}

All the data used in this paper are available within the article in graphical form
in the figures 1 to 9. If there is any difficulty in extracting them from the figures,
 the data can be provided on request from the author.

\end{document}